\documentclass[12pt]{article}
\usepackage[dvips]{graphicx,color}
\usepackage{amsmath,bm,calc}
\usepackage[psamsfonts]{amssymb}
\usepackage{longtable}
\begin{document}

\begin{titlepage}
\title{RPA calculations with Gaussian expansion method}

\author{H. Nakada\,$^a$, K. Mizuyama\,$^b$, M. Yamagami\,$^c$,
M. Matsuo\,$^d$\\
$^a$ \textit{Department of Physics, Graduate School of Science,
 Chiba University,}\\
\textit{Yayoi-cho 1-33, Inage, Chiba 263-8522, Japan}\\
$^b$ \textit{Department of Physics, University of Jyv\"{a}skyl\"{a},}\\
\textit{P. O. Box 35 (YFL), FI-40014, Jyv\"{a}skyl\"{a}, Finland}\\
$^c$ \textit{Department of Computer Science and Engineering,
 University of Aizu,}\\
\textit{Aizu-Wakamatsu, Fukushima 965-8580, Japan}\\
$^d$ \textit{Department of Physics, Faculty of Science,
 Niigata University,}\\
\textit{Niigata 950-2181, Japan}\\
}

\date{\today}
\maketitle
\thispagestyle{empty}

\begin{abstract}
The Gaussian expansion method (GEM) is applied
to calculations of the nuclear excitations
in the random-phase approximation (RPA).
We adopt the mass-independent basis-set
that is successful in the mean-field calculations.
The RPA results obtained by the GEM are compared
with those obtained by several other available methods in Ca isotopes,
by using a density-dependent contact interaction
along with the Woods-Saxon single-particle states.
It is confirmed that energies, transition strengths
and widths of their distribution are described
by the GEM with good precision,
for the $1^-$, $2^+$ and $3^-$ collective states.
The GEM is then applied to the self-consistent RPA calculations
with the finite-range Gogny D1S interaction.
The spurious center-of-mass motion is well separated
from the physical states in the $E1$ response,
and the energy-weighted sum rules for the isoscalar transitions
are fulfilled reasonably well.
Properties of low-energy transitions in $^{60}$Ca
are investigated in some detail.
\end{abstract}


\vspace*{3mm}\noindent
\end{titlepage}

\pagestyle{plain}

\section{Introduction}
\label{sec:intro}

As atomic nuclei far off the $\beta$-stability
exhibit exotic characters (\textit{e.g.} nucleon halos~\cite{ref:halo}
and new magic numbers~\cite{ref:magic}),
excitation properties of unstable nuclei come under interest
as well as their ground-state properties.
The new features of unstable nuclei,
such as the broad distribution of nucleons,
the exotic shell structure and the coupling to the continuum,
may give rise to new aspects of nuclear excitation.
In theoretical studies of unstable nuclei,
it is highly desired to handle those features efficiently.
It is expected that ground-state properties of many nuclei
can be investigated in the mean-field (MF) approximations,
although correlation effects are not necessarily negligible.
Among many efforts to develop new numerical methods
to study structure of unstable nuclei,
one of the authors (H.N.) proposed a method
for MF calculations~\cite{ref:NS02,ref:Nak06,ref:Nak08}
that is based on the Gaussian expansion method (GEM)~\cite{ref:GEM}.
The GEM is adaptable to the broad distribution of nucleons,
even with finite-range effective interactions
including semi-realistic nucleon-nucleon ($NN$)
interactions~\cite{ref:Nak03,ref:Nak08b}.
The results of the MF calculations suggest
that the continuum effects are properly taken into account
by the GEM~\cite{ref:Nak06}.
Moreover, we have found that the basis parameters in this method
are insensitive to nuclide, and thereby even a single set of bases
can be applied to wide mass range of nuclei~\cite{ref:Nak08}.
Because of these advantages,
it is of interest to use the GEM
for describing excitations of unstable nuclei.

The random-phase approximation (RPA) provides us
with a framework to treat
one-particle-one-hole ($1p$-$1h$) excitations of nuclei
in a consistent manner with the MF description of the ground state.
In this article we extensively apply the GEM to the RPA calculations.
The strength functions of the spin-independent transitions
are calculated in $^{40,48,60}$Ca.
We first use the single-particle (s.p.) states
in the Woods-Saxon potential
and adopt a density-dependent contact interaction
for the residual interaction.
For this schematic Hamiltonian,
reliable results are obtained by the continuum RPA~\cite{ref:SB75}.
We test the new method by comparing its results
with the continuum RPA results,
particularly for excitations of unstable nuclei.
There have been many RPA calculations
using the harmonic oscillator (HO) s.p. basis functions.
It is popular as well to implement RPA calculations
by assuming the box boundary for the s.p. states.
The new method is also compared with these methods.
The present method is then applied
to fully self-consistent Hartree-Fock (HF) plus RPA calculations,
by employing the finite-range Gogny interaction.
We shall check whether the spurious center-of-mass (c.m.) excitation
is well separated from the physical modes
and how well the energy-weighted sum rules are satisfied.
Properties of low-energy excitations in $^{60}$Ca
will also be discussed.

\section{Methods of calculations}
\label{sec:algorithm}

We shall consider the strength function,
\begin{equation}
 S(\omega) = \sum_\alpha \delta(\omega-\omega_\alpha)\,
  \big|\langle\alpha|\mathcal{O}|0\rangle\big|^2\,,
 \label{eq:S(E)}
\end{equation}
where $\mathcal{O}$ stands for the transition operator,
$|0\rangle$ the ground state,
$\omega_\alpha$ the excitation energy of the state $|\alpha\rangle$.
Below the particle-emission threshold we have discrete excited states.
The sum over $\alpha$ in Eq.~(\ref{eq:S(E)}) is replaced
by an energy integral for the continuum, if necessary.
The response function is defined by
\begin{equation}
 R(\omega) = \langle 0|\,
  \frac{\mathcal{O}^\dagger\mathcal{O}}{\omega-H+i\eta}\,|0\rangle\,,
 \label{eq:R(E)}
\end{equation}
with $\eta$ representing an infinitesimal positive number.
$R(\omega)$ is related to $S(\omega)$ through
\begin{equation}
 S(\omega) = -\frac{1}{\pi}\mathrm{Im}\big[R(\omega)\big]\,.
 \label{eq:S(E)-R(E)}
\end{equation}
By adding a finite imaginary number $i\gamma$ to the energy $\omega$,
we define
\begin{equation}
 R_\gamma(\omega) = \langle 0|\,
  \frac{\mathcal{O}^\dagger\mathcal{O}}{\omega-H+i\gamma}\,|0\rangle\,,
 \label{eq:R(E)'}
\end{equation}
from which $S(\omega)$ is smeared out as
\begin{equation}
 S_\gamma(\omega) = -\frac{1}{\pi}\mathrm{Im}\big[R_\gamma(\omega)\big]
  = \frac{1}{\pi}\sum_\alpha
  \frac{\gamma}{(\omega-\omega_\alpha)^2+\gamma^2}\,
  \big|\langle\alpha|\mathcal{O}|0\rangle\big|^2\,.
 \label{eq:S(E)'}
\end{equation}
For the $1p$-$1h$ excitations
$S(\omega)$ (or $S_\gamma(\omega)$) is calculated
from the transition amplitude $\langle\alpha|\mathcal{O}|0\rangle$
in the RPA,
or from $R(\omega)$ (or $R_\gamma(\omega)$) in the linear response theory.
It is well known that these two approaches
are equivalent to each other~\cite{ref:RS80}.
We therefore do not distinguish them in this paper.

We here discuss several methods
of calculating $S(\omega)$ or $S_\gamma(\omega)$ in the RPA.
If we take a set of basis functions,
the s.p. wave functions in the MF potential
are given by superposing the bases.
Solving the RPA equation that is represented by the s.p. states,
we obtain the excited state $|\alpha\rangle$
(to be more precise, forward and backward amplitudes
along with $\omega_\alpha$),
from which $S(\omega)$ (or $S_\gamma(\omega)$) is calculated.
In this type of methods, quality of the results is governed
by the s.p. basis functions.
In the RPA calculations of nuclei near the $\beta$-stability,
it has been customary to employ the HO basis-set.
However, it has been pointed out~\cite{ref:NS02}
that the HO set is impractical in handling the broad nucleon distribution
and the coupling to the continuum,
which could be important in unstable nuclei.
In the present paper we propose a method using the GEM basis-set,
which is comprised of multi-range Gaussian basis functions.
Adaptable to the energy-dependent asymptotics
of the s.p. wave functions including the oscillatory ones~\cite{ref:Nak06},
the GEM basis-set is expected to describe the $1p$-$1h$ excitations
appropriately, up to the continuum effects.

It is also popular to obtain the s.p. wave functions
by solving the MF equation in a discretized coordinate space.
To keep the number of the s.p. states finite,
one often confines the nucleons in a box
or imposes a periodic boundary condition.
In addition,
truncation with respect to the s.p. energies is necessary
in solving the RPA equation.
While this method is feasible for contact $NN$ interactions
(or for local densities and currents) such as the Skyrme interaction,
it is not easy to be applied
when a finite-range interaction is adopted\footnote{
It is mentioned that the Woods-Saxon basis functions,
which are obtained in a coordinate space,
have been used in Ref.~\cite{ref:GSB03}
for a quasiparticle RPA calculation with the Gogny interaction.}.

In the linear response theory
$R(\omega)$ (or $R_\gamma(\omega)$) is closely connected
to the s.p. Green's function.
The spectral representation,
\textit{i.e.} the expansion in terms of the s.p. states,
may be employed for the s.p. Green's function~\cite{ref:BT75}.
It is also possible to take account of
the exact form of the asymptotics of the s.p. Green's function
at each $\omega$,
as was argued and carried out in Ref.~\cite{ref:SB75}.
This method is often called `continuum RPA'.
It is an additional advantage of the continuum RPA
that we do not need truncation regarding the s.p. energies.
The strength function $S(\omega)$ or $S_\gamma(\omega)$
is straightforwardly calculated from $R(\omega)$ or $R_\gamma(\omega)$,
without constructing the excited state $|\alpha\rangle$ explicitly.
We note that the continuum RPA was extended to the quasiparticle RPA
in Refs.~\cite{ref:Mat01,ref:MMS09}.
Because it is implemented in a coordinate space,
the continuum RPA is suitable for contact $NN$ interactions.

There are other computational methods to implement the RPA calculations,
although we do not treat them in this article.
Recent developments include a method based on
the mixed representation of the RPA equation~\cite{ref:mix-rep},
in which the unoccupied s.p. states are handled
in a coordinate space~\cite{ref:LV68}.
The small amplitude limit of the time-dependent HF (TDHF) theory
derives the RPA~\cite{ref:RS80} as well.
Methods to compute the linear responses via the TDHF
have been explored~\cite{ref:NY05}.
One can carry out the TDHF calculations
either by using s.p. bases or in a coordinate space.
As a new approach based on the TDHF,
the finite-amplitude method has been invented~\cite{ref:NIY07}.

In all practical calculations in this paper,
we assume that the ground states of the nuclei
are well approximated by the spherical HF solutions.
We shall compare results of several methods for $^{40,48,60}$Ca
in Sec.~\ref{sec:contact}.
In the following of this section we present computational details
of the individual methods.
We will focus on the method using the GEM
in Sec.~\ref{sec:scRPA}.

\subsection{Gaussian expansion method\label{subsec:GEM-basis}}

In the present paper we newly introduce a method of the RPA calculations
using the Gaussian expansion method (GEM).
In the GEM, both the ground and the excited states are represented
by the s.p. basis functions having the following form:
\begin{eqnarray} \varphi_{\nu\ell jm}({\mathbf r})
&=& R_{\nu\ell j}(r)\,[Y^{(\ell)}(\hat{\mathbf r})
\chi_\sigma]^{(j)}_m\,; \nonumber\\
R_{\nu\ell j}(r) &=& \mathcal{N}_{\nu\ell j}\,r^\ell\exp(-\nu r^2)\,,
\label{eq:basis} \end{eqnarray}
where $Y^{(\ell)}(\hat{\mathbf r})$ expresses the spherical harmonics
and $\chi_\sigma$ the spin wave function.
We drop the isospin index without confusion.
The range parameter of the Gaussian basis function $\nu$,
which is also used as an index of the bases,
can be complex ($\nu=\nu_\mathrm{r}+i\nu_\mathrm{i}$)~\cite{ref:GEM}.
The constant $\mathcal{N}_{\nu\ell j}$ is determined by
\begin{equation} \mathcal{N}_{\nu\ell j}
= \frac{2^{\ell+\frac{7}{4}}}{\pi^\frac{1}{4}\sqrt{(2\ell+1)!!}}\,
 \,\nu_\mathrm{r}^\frac{2\ell+3}{4},
\end{equation}
so as for $\langle\varphi_{\nu\ell jm}|\varphi_{\nu\ell jm}
\rangle$ to be unity.
We take the set comprised of all the bases of
\begin{equation}
\nu_\mathrm{r}=\nu_0\,b^{-2k}\,,\quad
\left\{\begin{array}{ll}\nu_\mathrm{i}=0 & (k=0,1,\cdots,5)\\
{\displaystyle\frac{\nu_\mathrm{i}}{\nu_\mathrm{r}}
=\pm\frac{\pi}{2}} & (k=0,1,2)\end{array}\right.\,,
 \label{eq:basis-param}
\end{equation}
with $\nu_0=(2.40\,\mathrm{fm})^{-2}$ and $b=1.25$
($6$ real functions and $6$ complex functions),
irrespectively of $(\ell,j)$.
With this basis-set we can describe ground states of many nuclei
in a wide mass range from $^{16}$O to $^{208}$Pb~\cite{ref:Nak08}.
We first solve a MF equation in the s.p. space
spanned by the above bases.
We note that, since the GEM bases are not orthogonal to one another,
the MF equation leads to a generalized eigenvalue problem.
The MF solution determines the s.p. states,
by which we represent the RPA equation.
We then solve the RPA equation,
following Sec.~8.4.4 of Ref.~\cite{ref:RS80}.

\subsection{Method of quasi-HO basis functions\label{subsec:qHO-basis}}

The HO basis-set has been employed
for many MF and RPA calculations so far.
The radial part of the spherical HO basis function
has the form of a Gaussian
multiplied by the associated Laguerre's polynomial.
As mentioned in Ref.~\cite{ref:NS02},
we can produce a basis-set that is equivalent to the HO set
by extending the basis functions of Eq.~(\ref{eq:basis});
$R(r)\propto r^{\ell+2p}\exp(-\nu r^2)$ with an integer $p$,
and $\nu$ restricted to a single real value.
While the individual basis functions
are not the same as the HO functions,
the set comprised of them is equivalent to the HO basis-set
of $\{L_p^{(\ell+\frac{1}{2})}(\sqrt{2\nu}r)\,
r^\ell\exp(-\nu r^2); p=0,1,2,\cdots\}$ for each $(\ell,j)$,
where $L_p^{(\alpha)}(x)$ is the associated Laguerre's polynomial,
as revealed if we apply the Gram-Schmidt orthogonalization.
An analogous basis-set is produced
from the basis functions of Eq.~(\ref{eq:basis}),
by taking $\nu=\nu_0\,b^{-2k}$ ($k=0,1,2,\cdots$)
with real $\nu_0$ and $b$ close to unity.
Indeed, if we take the $b=1$ limit
after orthogonalizing to the $k=0$ basis
$R_{\nu_0 \ell j}(r)\propto r^\ell\exp(-\nu_0 r^2)
=L_0^{(\ell+\frac{1}{2})}(\sqrt{2\nu_0}r)\,
r^\ell\exp(-\nu_0 r^2)$,
the $k=1$ basis yields the HO basis of
$L_1^{(\ell+\frac{1}{2})}(\sqrt{2\nu_0}r)\,r^\ell\exp(-\nu_0 r^2)$.
It is easy to show that the basis proportional to
$L_k^{(\ell+\frac{1}{2})}(\sqrt{2\nu_0}r)\,r^\ell\exp(-\nu_0 r^2)$
is produced by successive orthogonalization,
and thereby we obtain a set equivalent to the HO basis-set.
Obviously $2k+\ell$ corresponds to
the number of the oscillator quanta $N_\mathrm{osc}$ in the HO bases.
In practice, we adopt a set of the basis functions of Eq.~(\ref{eq:basis})
with $b=1.05$ and $\nu_0$ determined from
$\omega_\mathrm{osc}=41.2A^{-1/3}\,\mathrm{MeV}$
through $\nu_0=\sqrt{2/M\omega_\mathrm{osc}}$.
This set will be called `quasi-HO' basis-set hereafter.
The space spanned by the bases satisfying $2k+\ell\leq 11$ is taken
in the calculations in Sec.~\ref{sec:contact}.
With this basis-set we solve the MF and RPA equations
in the same manner as in Subsec.~\ref{subsec:GEM-basis}.

\subsection{Coordinate-space method with box boundary
  \label{subsec:box-basis}}

If the MF is local,
it is not difficult to solve the MF equation in a coordinate space.
Because we assume the spherical symmetry,
it is sufficient to consider one-dimensional coordinate space
of the radial variable.
This radial coordinate is discretized by a mesh
whose size is denoted by $h$.
We impose a box boundary condition
that the s.p. wave functions should vanish at $r_\mathrm{max}$.
The MF equation is solved in the coordinate space thus determined.
In Sec.~\ref{sec:contact} we adopt $h=0.2\,\mathrm{fm}$
and $r_\mathrm{max}=20\,\mathrm{fm}$,
confirming the convergence in the MF calculations.

The RPA Hamiltonian is represented by the s.p. states
obtained in the MF calculation,
by cutting off the unperturbed excitation energy
at $\mathit{\Delta}\varepsilon_\mathrm{cut}$.
In practical calculations in Sec.~\ref{sec:contact},
we do not solve the eigenvalue problem defined by the RPA equation,
but we instead compute $R_\gamma(\omega)$
by solving the Bethe-Salpeter equation
in the spectral representation~~\cite{ref:BT75}
(using the s.p. states and $\mathit{\Delta}\varepsilon_\mathrm{cut}$).
The strength function $S_\gamma(\omega)$ is then obtained
via Eq.~(\ref{eq:S(E)-R(E)}) (or (\ref{eq:S(E)'})).
For certain transition modes the RPA results are slow to converge
for increasing $\mathit{\Delta}\varepsilon_\mathrm{cut}$~\cite{ref:BG77}.
We have confirmed that $S_\gamma(\omega)$ at $\omega<30\,\mathrm{MeV}$
has no visible difference
between $\mathit{\Delta}\varepsilon_\mathrm{cut}=150\,\mathrm{MeV}$
and $200\,\mathrm{MeV}$
for all the modes under consideration.

\subsection{Continuum RPA\label{subsec:cRPA}}

In the continuum RPA~\cite{ref:SB75},
the strength function is calculated from the response function.
However, the response function is constructed
from the s.p. orbits of holes and the s.p. Green's function
for the excited particles,
unlike the spectral representation.
The asymptotic form of the s.p. Green's function at large $r$,
which depends on $\omega$, is known for each partial wave.
For each $\omega$, the inner part of the s.p. Green's function
is continuated to the proper asymptotic form at a sufficiently large $r$.
Hence the calculation is free both from the boundary at $r_\mathrm{max}$
and from the energy cut-off like $\mathit{\Delta}\varepsilon_\mathrm{cut}$.
For a relatively simple interaction as used in Sec.~\ref{sec:contact},
we can obtain `exact' transition strength functions
(which are exact within the RPA),
as long as the convergence with respect to $h$ is reached.
In the continuum RPA calculations in Sec.~\ref{sec:contact},
we follow the method shown in Ref.~\cite{ref:Mat01},
except that we do not need integration in the complex energy plane
because we do not have pair correlations in the ground state.
The mesh size $h$ is taken to be $0.2\,\mathrm{fm}$
as in the box-boundary calculation,
for which the convergence has been confirmed.

Whereas the continuum RPA calculations are implemented
with specifying $\omega$,
it is not easy to pinpoint the excitation energies
of the discrete states.
To avoid missing the discrete levels,
we calculate $S_\gamma(\omega)$ rather than $S(\omega)$,
taking $\gamma=0.2\,\mathrm{MeV}$.
The excitation energy and the transition strength of each discrete state
is extracted by fitting $S_\gamma(\omega)$ around the peak
to a Lorentzian.
Correspondingly, we compute $S_\gamma(\omega)$
with $\gamma=0.2\,\mathrm{MeV}$ in the other methods,
by using the imaginary energy $\omega+i\gamma$
in the method of Subsec.~\ref{subsec:box-basis}
and by smearing out $S(\omega)$ in those of
Subsecs.~\ref{subsec:GEM-basis} and \ref{subsec:qHO-basis}.

\section{Comparison of methods for contact force\label{sec:contact}}

In this section numerical results of the four methods
explained in the preceding section
\ref{subsec:GEM-basis}--\ref{subsec:cRPA}
are compared.
For this purpose,
we take the nuclear MF of the Woods-Saxon potential
to which the Coulomb potential is added,
as adopted in Ref.~\cite{ref:SB75}:
\begin{equation}
 U(\mathbf{r}) = \left(1-0.67\frac{N-Z}{A}\tau_z\right)
  \left(U_0\,f(r) + U_{\ell s}\,\mbox{\mathversion{bold}$\ell$}\cdot
   \mathbf{s}\,\frac{1}{r} \frac{d}{dr}f(r)\right)
  + \frac{1}{2}(1-\tau_z)\,U_C(r)\,,
  \label{eq:WS}
\end{equation}
where $f(r)=1/[1+\exp((r-R)/a)]$,
$U_0=-58\,\mathrm{MeV}$, $U_{\ell s}=30\,\mathrm{MeV}\,\mathrm{fm}^2$,
$R=1.20(A-1)^{1/3}\,\mathrm{fm}$, $a=0.65\,\mathrm{fm}$,
$\tau_z=+1$ ($-1$) for a neutron (a proton),
and $U_C(r)$ is the Coulomb potential
produced by the uniform charge distribution of $(Z-1)$ protons
in the sphere of radius $R$.
For the residual interaction,
we assume the density-dependent contact force as
\begin{equation}
 \hat{v}_\mathrm{res} = f\big[t_0 (1+x_0 P_\sigma)\,
  \delta(\mathbf{r}_1-\mathbf{r}_2)
  + \frac{1}{6}t_3 (1+x_3 P_\sigma)\,\rho(\mathbf{r}_1)\,
  \delta(\mathbf{r}_1-\mathbf{r}_2)\big]\,,
  \label{eq:SB}
\end{equation}
with $t_0=-1100\,\mathrm{MeV}\,\mathrm{fm}^3$, $x_0=0.5$,
$t_3=16000\,\mathrm{MeV}\,\mathrm{fm}^6$ and $x_3=1$~\cite{ref:SB75}.
$\rho(\mathbf{r})$ denotes the nucleon density,
and $P_\sigma$ the spin exchange operator.
It will be stated below how to fix the overall factor $f$.
The spin densities arising from $\hat{v}_\mathrm{res}$ are ignored
for the sake of simplicity.

We shall consider transitions carried by the one-body operator,
\begin{equation}
 \mathcal{O}^{(\lambda,\tau)}
 = \sum_i r_i^\lambda Y^{(\lambda)}(\hat{\mathbf{r}}_i)
 \left\{\begin{array}{cl} 1&\mbox{(for $\tau=0$)}\\
  \tau_z&\mbox{(for $\tau=1$)} \end{array} \right.,
  \label{eq:op23}
\end{equation}
with $\lambda=1,2,3$, where $i$ denotes the index of nucleons,
for the $^{40}$Ca, $^{48}$Ca and $^{60}$Ca nuclei.
The $(\lambda=1,\tau=0)$ mode corresponds to the spurious c.m. motion.
Subtracting its contribution,
we obtain a modified operator for the $(\lambda=1,\tau=1)$ mode,
\begin{equation}
 \tilde{\mathcal{O}}^{(\lambda=1,\tau=1)}
 = \frac{2Z}{A}\sum_{i\in n} r_i Y^{(1)}(\hat{\mathbf{r}}_i)
  -\frac{2N}{A}\sum_{i\in p} r_i Y^{(1)}(\hat{\mathbf{r}}_i)\,,
  \label{eq:op1}
\end{equation}
which is proportional to the $E1$ operator with the c.m. correction.
The expression $i\in n$ ($i\in p$) indicates
that the sum includes the $i$-th nucleon if it is a neutron (proton).
The strength functions $S^{(\lambda,\tau)}$
and $S_\gamma^{(\lambda,\tau)}$ are defined
by substituting $\mathcal{O}^{(\lambda,\tau)}$
in Eqs.~(\ref{eq:S(E)},\ref{eq:S(E)'}).
For the $(\lambda=1,\tau=1)$ mode
we can define $\tilde{S}^{(\lambda=1,\tau=1)}$
and $\tilde{S}_\gamma^{(\lambda=1,\tau=1)}$
from $\tilde{\mathcal{O}}^{(\lambda=1,\tau=1)}$.
The renormalization parameter $f$ in Eq.~(\ref{eq:SB}) is determined
so that the spurious c.m. state should have
zero excitation energy~\cite{ref:SB75}, for individual cases.
We tabulate the $f$ values thus determined in Table~\ref{tab:f-val}.
Although there is no consistency between the MF
and the residual interaction,
this set of the MF potential and the interaction is suitable
for assessing the numerical methods.

\begin{table}
\begin{center}
\caption{Adopted values of $f$,
 the renormalization factor of the residual interaction.
 The label `Cont.' (`Box') indicates the case
 of the continuum RPA (the box-boundary method).
\label{tab:f-val}}
\begin{tabular}{crrrr}
\hline\hline
nuclide & Cont. & Box~~& quasi-HO & GEM \\
 \hline
$^{40}$Ca &\quad$0.691$ &\quad$0.785$ &\quad$0.715$ &\quad$0.700$ \\
$^{48}$Ca & $0.711$ & $0.808$ & $0.750$ & $0.720$ \\
$^{60}$Ca & $0.758$ & $0.842$ & $0.796$ & $0.764$ \\
\hline\hline
\end{tabular}
\end{center}
\end{table}

\begin{figure}
%\centerline{\includegraphics[scale=0.85]{delta/E1IV_Ca.eps}}
\centerline{\includegraphics[scale=0.85]{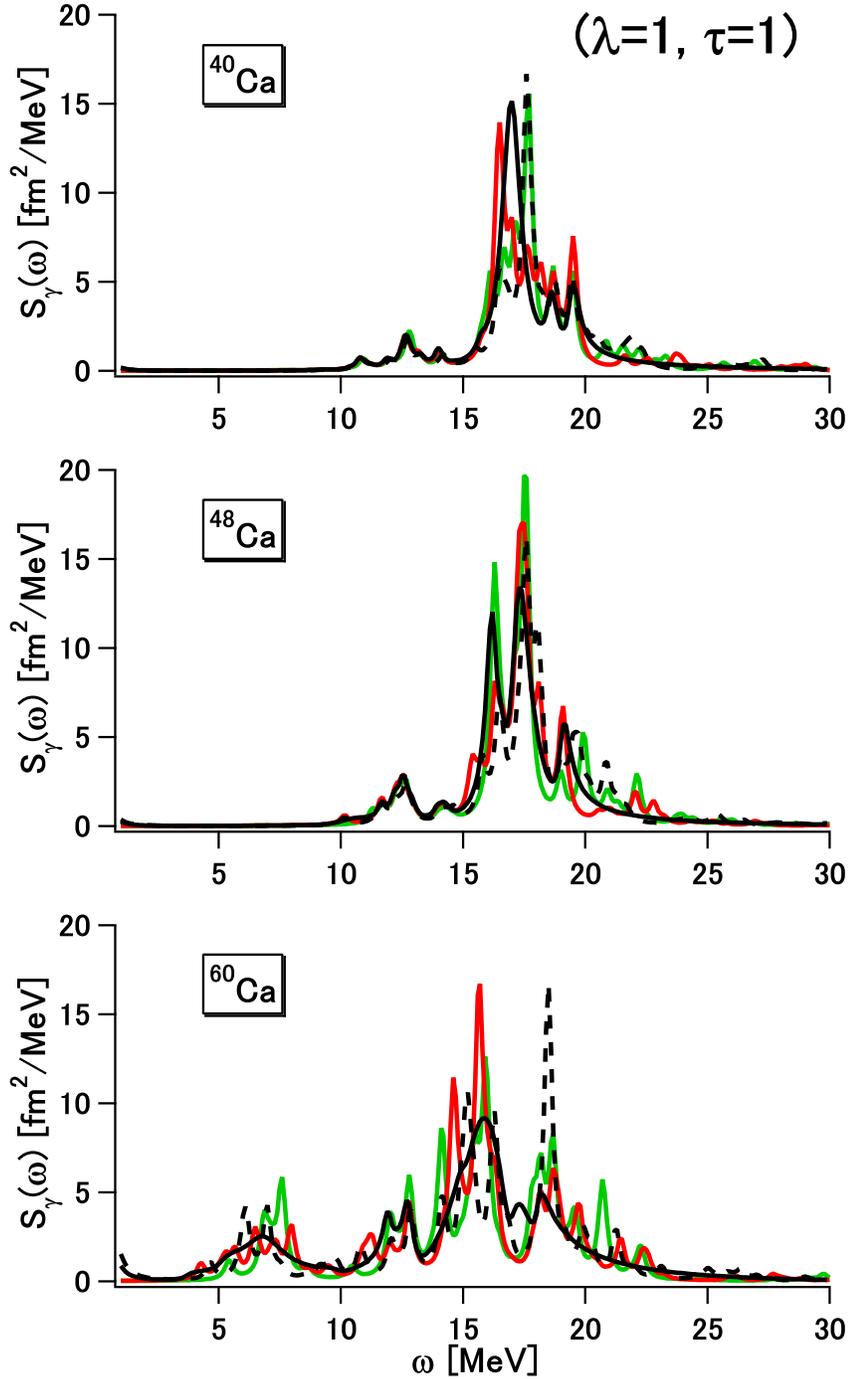}}
\vspace{2mm}
\caption{$\tilde{S}_\gamma^{(\lambda=1,\tau=1)}(\omega)$
 in $^{40,48,60}$Ca, by the RPA calculations
 using the mean field of Eq.~(\protect\ref{eq:WS})
 and the residual interaction of Eq.~(\protect\ref{eq:SB}).
 Results of the continuum RPA, the box-boundary method,
 the method employing the quasi-HO bases
 and the GEM are represented by the black solid,
 the black dashed, the green solid and the red solid lines,
 respectively.
 We take $\gamma=0.2\,\mathrm{MeV}$ for all the calculations.
\label{fig:E1IV_comp}}
\end{figure}

\begin{figure}
%\centerline{\includegraphics[scale=0.85]{delta/E2IS_Ca.eps}}
\centerline{\includegraphics[scale=0.85]{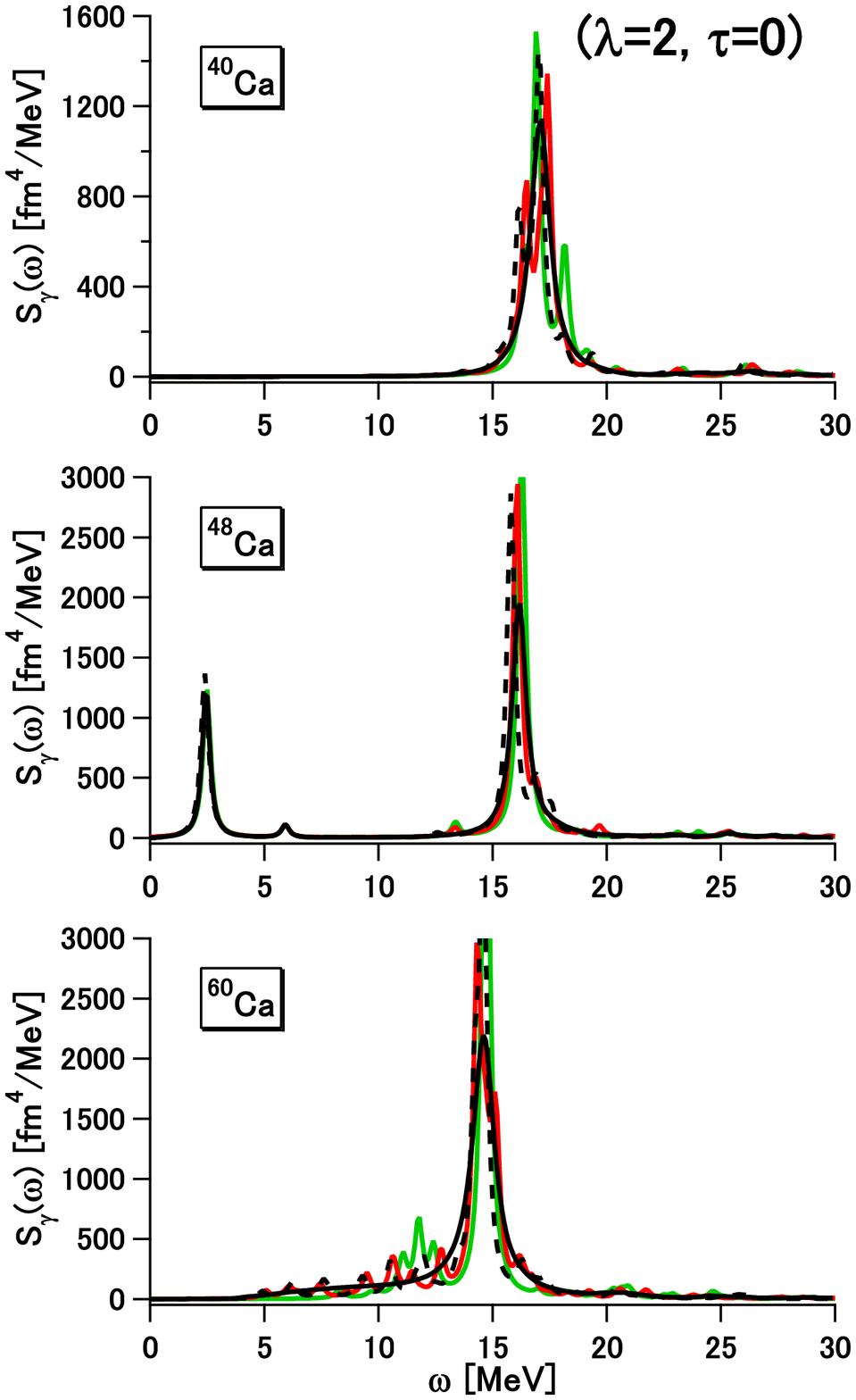}}
\vspace{2mm}
\caption{$S_\gamma^{(\lambda=2,\tau=0)}(\omega)$ in $^{40,48,60}$Ca.
 See Fig.~\protect\ref{fig:E1IV_comp} for conventions.
\label{fig:E2IS_comp}}
\end{figure}

\begin{figure}
%\centerline{\includegraphics[scale=0.85]{delta/E2IV_Ca.eps}}
\centerline{\includegraphics[scale=0.85]{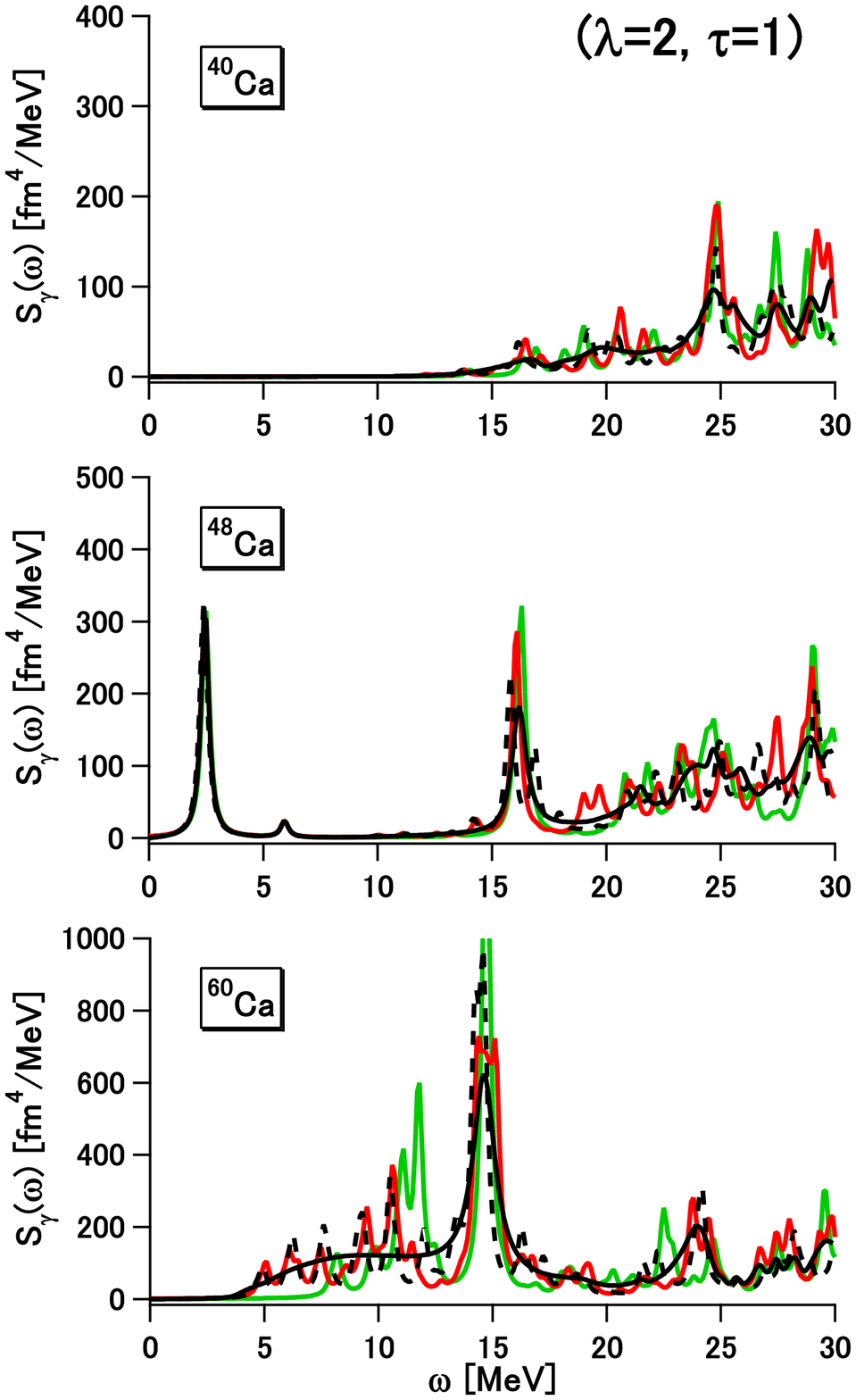}}
\vspace{2mm}
\caption{$S_\gamma^{(\lambda=2,\tau=1)}(\omega)$ in $^{40,48,60}$Ca.
 See Fig.~\protect\ref{fig:E1IV_comp} for conventions.
\label{fig:E2IV_comp}}
\end{figure}

\begin{figure}
%\centerline{\includegraphics[scale=0.85]{delta/E3IS_Ca.eps}}
\centerline{\includegraphics[scale=0.85]{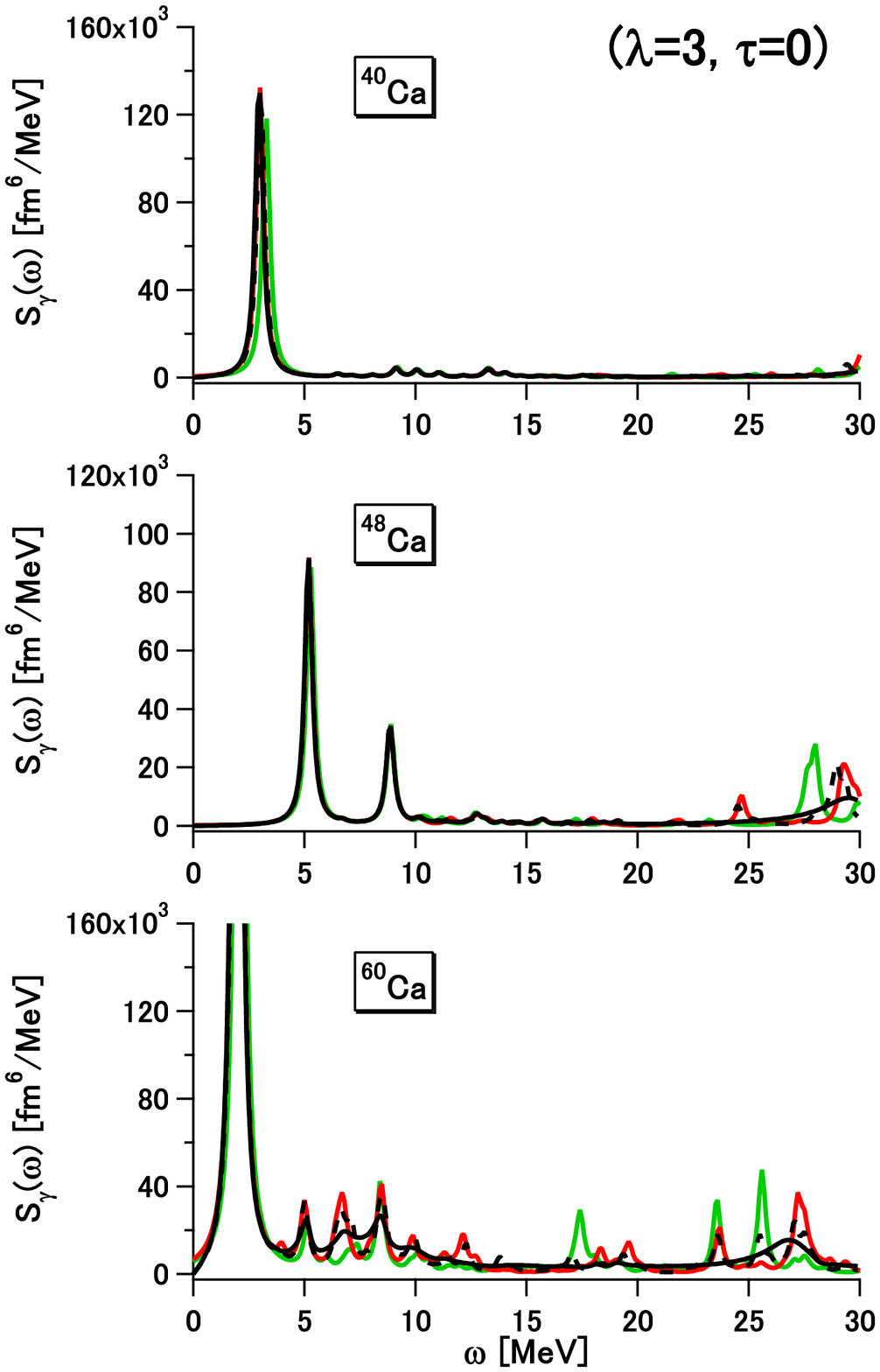}}
\vspace{2mm}
\caption{$S_\gamma^{(\lambda=3,\tau=0)}(\omega)$ in $^{40,48,60}$Ca.
 See Fig.~\protect\ref{fig:E1IV_comp} for conventions.
\label{fig:E3IS_comp}}
\end{figure}

\begin{figure}
%\centerline{\includegraphics[scale=0.85]{delta/E3IV_Ca.eps}}
\centerline{\includegraphics[scale=0.85]{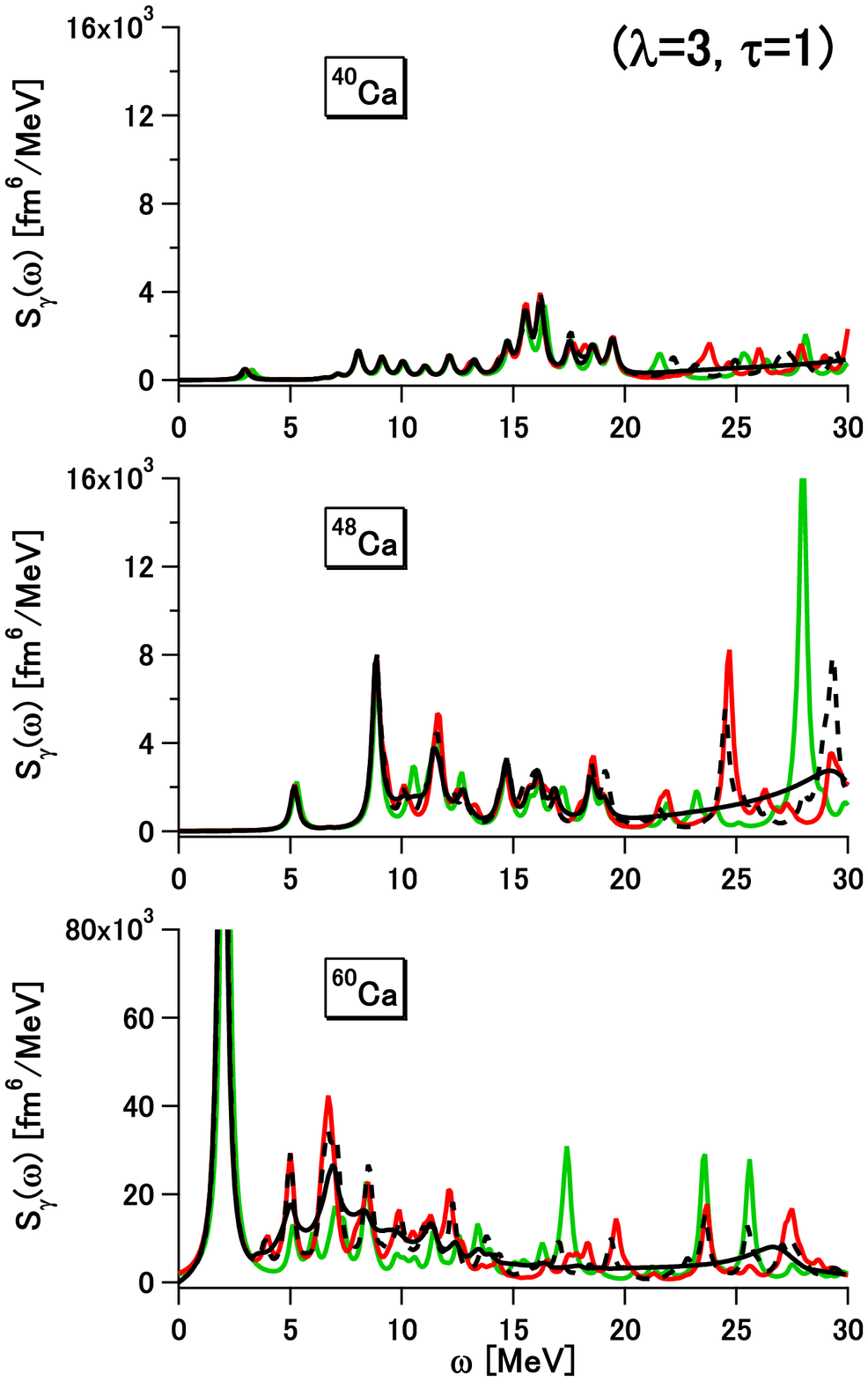}}
\vspace{2mm}
\caption{$S_\gamma^{(\lambda=3,\tau=1)}(\omega)$ in $^{40,48,60}$Ca.
 See Fig.~\protect\ref{fig:E1IV_comp} for conventions.
\label{fig:E3IV_comp}}
\end{figure}

In Figs.~\ref{fig:E1IV_comp}--\ref{fig:E3IV_comp}
we display the strength functions
$S_\gamma^{(\lambda,\tau)}(\omega)$.
For the calculations assuming the box boundary
shown in Subsec.~\ref{subsec:box-basis},
we take $\mathit{\Delta}\varepsilon_\mathrm{cut}=50\,\mathrm{MeV}$,
whose results are not qualitatively different
from those of $\mathit{\Delta}\varepsilon_\mathrm{cut}=200\,\mathrm{MeV}$.

If we estimate from the s.p. energies,
the particle threshold lies at $\omega=8.7\,\mathrm{MeV}$
for $^{40}$Ca and $^{48}$Ca,
while at $3.4\,\mathrm{MeV}$ for $^{60}$Ca.
In all the three nuclei we find $\lambda=3$ peaks at low $\omega$
that correspond to discrete states.
For $^{48}$Ca we also find low-lying discrete $2^+$ states
because of the shell structure.
Although the $8.86\,\mathrm{MeV}$ state
in the $\lambda=3$ excitation of $^{48}$Ca
is located just above the neutron threshold,
it behaves like a discrete state.
For these discrete states
we list the excitation energies $\omega_\alpha$
and the transition strengths
\begin{equation}
 B^{(\lambda,\tau)}_\alpha
  = \big| \langle\alpha||\mathcal{O}^{(\lambda,\tau)} ||0\rangle \big|^2
  = \int_{\omega_\alpha-\eta}^{\omega_\alpha+\eta}
  S^{(\lambda,\tau)}(\omega)\,d\omega\,,
\end{equation}
in Table~\ref{tab:discrete}.
The low-lying $3^-$ state is highly collective in any of the three nuclei.
For this state
the convergence for $\mathit{\Delta}\varepsilon_\mathrm{cut}$
is quite slow in the calculation under the box boundary condition,
and we show the results
with $\mathit{\Delta}\varepsilon_\mathrm{cut}=200\,\mathrm{MeV}$
in Table~\ref{tab:discrete}.
Note that the $S_\gamma^{(\lambda,\tau)}(\omega)$ graphs
in Figs.~\ref{fig:E3IS_comp} and \ref{fig:E3IV_comp} do not much change
even if we take $\mathit{\Delta}\varepsilon_\mathrm{cut}=200\,\mathrm{MeV}$.

\hspace*{-1.0cm}\rotatebox{90}{\begin{minipage}{22cm}
\begin{longtable}{cccrrrrrrrr}
%\begin{center}
\caption{Excitation energy $\omega_\alpha$ (MeV)
 and transition strength $B^{(\lambda,\tau)}_\alpha$ (fm$^{2\lambda}$)
 for each discrete state,
 obtained by the RPA calculations
 employing the mean field of Eq.~(\protect\ref{eq:WS})
 and the residual interaction of Eq.~(\protect\ref{eq:SB}).
 Results of the continuum RPA, the box-boundary method,
 the method using the quasi-HO bases and the GEM are compared.
\label{tab:discrete}}\\
%\begin{tabular}{cccrrrrrrrr}
\hline\hline
nuclide &&& \multicolumn{2}{c}{Cont.} & \multicolumn{2}{c}{Box} &
 \multicolumn{2}{c}{quasi-HO} & \multicolumn{2}{c}{GEM} \\
 & $\lambda$ & $\tau$ &
 \multicolumn{1}{c}{\quad$\omega_\alpha$} &
 \multicolumn{1}{c}{$B^{(\lambda,\tau)}_\alpha$} &
 \multicolumn{1}{c}{\quad$\omega_\alpha$} &
 \multicolumn{1}{c}{$B^{(\lambda,\tau)}_\alpha$} &
 \multicolumn{1}{c}{\quad$\omega_\alpha$} &
 \multicolumn{1}{c}{$B^{(\lambda,\tau)}_\alpha$} &
 \multicolumn{1}{c}{\quad$\omega_\alpha$} &
 \multicolumn{1}{c}{$B^{(\lambda,\tau)}_\alpha$} \\
 \hline
$^{40}$Ca & $3$ & $0$ &\quad $2.96$ & $8.49\times 10^4$ &\quad
 $3.04$ & $8.20\times 10^4$ &\quad $3.28$ & $7.51\times 10^4$ &\quad
 $3.00$ & $8.32\times 10^4$ \\
 && $1$ && $3.29\times 10^2$ && $3.25\times 10^2$ &&
 $2.97\times 10^2$ && $3.20\times 10^2$ \\
$^{48}$Ca & $2$ & $0$ & $2.45$ & $7.88\times 10^2$ &
 $2.45$ & $7.87\times 10^2$ & $2.49$ & $7.78\times 10^2$ &
 $2.45$ & $7.89\times 10^2$ \\
 && $1$ && $2.03\times 10^2$ && $2.02\times 10^2$ &&
 $1.99\times 10^2$ && $2.03\times 10^2$ \\
 & $2$ & $0$ & $5.92$ & $7.38\times 10^1$ &
 $5.92$ & $7.38\times 10^1$ & $5.93$ & $6.68\times 10^1$ &
 $5.92$ & $6.85\times 10^1$ \\
 && $1$ && $1.57\times 10^1$ && $1.56\times 10^1$ &&
 $1.42\times 10^1$ && $1.43\times 10^1$ \\
 & $3$ & $0$ & $5.18$ & $5.82\times 10^4$ &
 $5.20$ & $5.77\times 10^4$ & $5.27$ & $5.66\times 10^4$ &
 $5.19$ & $5.77\times 10^4$ \\
 && $1$ && $1.33\times 10^3$ && $1.36\times 10^3$ &&
 $1.41\times 10^3$ && $1.31\times 10^3$ \\
 & $3$ & $0$ & $8.86$ & $2.22\times 10^4$ &
 $8.86$ & $2.24\times 10^4$ & $8.89$ & $2.17\times 10^4$ &
 $8.86$ & $2.17\times 10^4$ \\
 && $1$ && $5.35\times 10^3$ && $5.48\times 10^3$ &&
 $4.68\times 10^3$ && $4.81\times 10^3$ \\
$^{60}$Ca & $3$ & $0$ & $1.96$ & $4.07\times 10^5$ &
 $1.98$ & $4.02\times 10^5$ & $2.09$ & $3.91\times 10^5$ &
 $1.97$ & $4.06\times 10^5$ \\
 && $1$ && $1.15\times 10^5$ && $1.15\times 10^5$ &&
 $1.17\times 10^5$ && $1.15\times 10^5$ \\
\hline\hline
%\end{tabular}
%\end{center}
\end{longtable}
\end{minipage}}

By comparing with the results of the continuum RPA,
we find that the maximum deviation of $\omega_\alpha$
is $0.08$, $0.32$ and $0.04\,\mathrm{MeV}$
in the box-boundary, the quasi-HO and the GEM results,
respectively.
The relative error of $B_\alpha^{(\lambda,\tau)}$
is $3\%$ in the box-boundary, $13\%$ in the quasi-HO
and $10\%$ in the GEM results, at maximum.
Thus, as long as the discrete states are concerned,
the GEM provides precision of $\sim 0.05\,\mathrm{MeV}$
for $\omega_\alpha$ and $\sim 10\%$ for $B_\alpha$.
The precision of the GEM for $\omega_\alpha$ is
slightly better than that of the box-boundary calculations,
and substantially better than that of the calculations
using the quasi-HO bases.
Although the precision might look worse than the box-boundary method
for $B_\alpha$,
the strong transitions are described with good precision.
It is noticed that the quasi-HO basis-set gives rise to sizable errors
for the lowest-lying $3^-$ states.

Above the particle-emission threshold,
the continuum RPA gives smooth $S(\omega)$
(\textit{i.e.} the $\gamma\rightarrow 0$ limit of $S_\gamma(\omega)$),
whereas in the other methods $S(\omega)$ is still represented
by a sum of the delta functions.
In Figs.~\ref{fig:E1IV_comp}--\ref{fig:E3IV_comp}
the strength functions are smoothed
by the width parameter $\gamma\,(=0.2\,\mathrm{MeV})$.
It is a clear advantage of the continuum RPA
that the smooth behavior of $S(\omega)$ in the continuum
is automatically taken into account,
and could be important in investigating certain aspects of transitions,
\textit{e.g.} the soft dipole responses
near the threshold~\cite{ref:Mat01,ref:MMS05}.
However, at higher $\omega$
the strength functions should further be smeared
due to coupling to the $2p$-$2h$ degrees of freedom,
which is not incorporated in the usual RPA.
Therefore gross properties of the strength distribution
will be more important than its fine structure,
when we view the RPA results.
In Table~\ref{tab:summed},
we tabulate the transition strength, the average energy and
its standard deviation in an energy domain
$D=\{\omega;\,\omega_D^\mathrm{min}<\omega<\omega_D^\mathrm{max}\}$;
\begin{eqnarray}
 B^{(\lambda,\tau)}_D &=&
  \int_D S_\gamma^{(\lambda,\tau)}(\omega)\,d\omega\,,\nonumber\\
 \bar{\omega}^{(\lambda,\tau)}_D &=& \left.
  \int_D \omega\,S_\gamma^{(\lambda,\tau)}(\omega)\,d\omega
  \right/ B^{(\lambda,\tau)}_D\,,\nonumber\\
 \sigma^{(\lambda,\tau)}_D &=& \left[\left.
  \int_D (\omega-\bar{E_x})^2\,S_\gamma^{(\lambda,\tau)}(\omega)\,
  d\omega \right/ B^{(\lambda,\tau)}_D \right]^{1/2}\,.
 \label{eq:B-dist}
\end{eqnarray}
The domain $D$ is chosen
so as to cover the broad resonance-like structure
of $S^{(\lambda,\tau)}(\omega)$.
For the $(\lambda=1,\tau=1)$ mode,
we use $\tilde{S}_\gamma^{(\lambda=1,\tau=1)}(\omega)$
instead of $S_\gamma^{(\lambda=1,\tau=1)}(\omega)$
as before.

\newpage\vspace*{1.0cm}\hspace*{-1.0cm}
\rotatebox{90}{\begin{minipage}{22cm}
\begin{small}
\begin{longtable}{ccccrrrrrrrrrrrr}
%\begin{center}
\caption{\normalsize
 Comparison of $B^{(\lambda,\tau)}_D$ ($\mathrm{fm}^{2\lambda}$),
 $\bar{\omega}^{(\lambda,\tau)}_D$ ($\mathrm{MeV}$)
 and $\sigma^{(\lambda,\tau)}_D$ ($\mathrm{MeV}$)
 in a certain energy range $D$,
 which is specified by $\omega_D^\mathrm{min}$
 and $\omega_D^\mathrm{max}$ ($\mathrm{MeV}$).
\label{tab:summed}}\\
%\begin{tabular}{ccccrrrrrrrrrrrr}
\hline\hline
nuclide &&&&
 \multicolumn{3}{c}{Cont.} & \multicolumn{3}{c}{Box} &
 \multicolumn{3}{c}{quasi-HO} & \multicolumn{3}{c}{GEM} \\
 & $\lambda$ & $\tau$ & $D$ &
 \multicolumn{1}{c}{\quad$B^{(\lambda,\tau)}_D$} &
 \multicolumn{1}{c}{$\bar{\omega}^{(\lambda,\tau)}_D$} &
 \multicolumn{1}{c}{$\sigma^{(\lambda,\tau)}_D$} &
 \multicolumn{1}{c}{\quad$B^{(\lambda,\tau)}_D$} &
 \multicolumn{1}{c}{$\bar{\omega}^{(\lambda,\tau)}_D$} &
 \multicolumn{1}{c}{$\sigma^{(\lambda,\tau)}_D$} &
 \multicolumn{1}{c}{\quad$B^{(\lambda,\tau)}_D$} &
 \multicolumn{1}{c}{$\bar{\omega}^{(\lambda,\tau)}_D$} &
 \multicolumn{1}{c}{$\sigma^{(\lambda,\tau)}_D$} &
 \multicolumn{1}{c}{\quad$B^{(\lambda,\tau)}_D$} &
 \multicolumn{1}{c}{$\bar{\omega}^{(\lambda,\tau)}_D$} &
 \multicolumn{1}{c}{$\sigma^{(\lambda,\tau)}_D$} \\
 \hline
$^{40}$Ca & $1$ & $1$ & $10.0-25.0$ &
 \quad $7.78\times 10^0$ & $17.35$ & $2.38$ &
 \quad $7.44\times 10^0$ & $17.85$ & $2.40$ &
 \quad $7.72\times 10^0$ & $17.54$ & $2.38$ &
 \quad $7.82\times 10^0$ & $17.43$ & $2.43$ \\
 & $2$ & $0$ & $12.0-22.0$ & $1.66\times 10^3$ & $17.12$ & $1.10$ &
 $1.67\times 10^3$ & $16.88$ & $1.14$ &
 $1.63\times 10^3$ & $17.23$ & $1.12$ &
 $1.66\times 10^3$ & $17.09$ & $1.10$ \\
 && $1$ & $12.0-30.0$ & $6.61\times 10^2$ & $24.49$ & $3.98$ &
 $5.95\times 10^2$ & $24.29$ & $3.99$ &
 $6.43\times 10^2$ & $24.65$ & $3.72$ &
 $6.88\times 10^2$ & $24.73$ & $4.06$ \\
$^{48}$Ca & $1$ & $1$ & $10.0-25.0$ &
 $9.33\times 10^0$ & $17.06$ & $2.45$ &
 $8.92\times 10^0$ & $17.61$ & $2.47$ &
 $9.29\times 10^0$ & $17.37$ & $2.56$ &
 $9.40\times 10^0$ & $17.15$ & $2.49$ \\
 & $2$ & $0$ & $12.0-22.0$ & $2.29\times 10^3$ & $16.32$ & $1.16$ &
 $2.28\times 10^3$ & $16.13$ & $1.13$ &
 $2.26\times 10^3$ & $16.30$ & $1.07$ &
 $2.29\times 10^3$ & $16.30$ & $1.19$ \\
 && $1$ & $12.0-30.0$ & $1.08\times 10^3$ & $23.36$ & $4.67$ &
 $9.90\times 10^2$ & $23.09$ & $4.80$ &
 $1.04\times 10^3$ & $23.40$ & $4.70$ &
 $1.09\times 10^3$ & $23.43$ & $4.70$ \\
 & $3$ & $1$ & $10.0-22.0$ & $1.71\times 10^4$ & $15.08$ & $3.22$ &
 $1.62\times 10^4$ & $15.30$ & $3.25$ &
 $1.55\times 10^4$ & $14.65$ & $3.20$ &
 $1.70\times 10^4$ & $15.11$ & $3.27$ \\
$^{60}$Ca & $1$ & $1$ & $~2.0-10.0$ &
 $2.11\times 10^0$ & $6.75$ & $1.54$ &
 $1.90\times 10^0$ & $6.71$ & $1.55$ &
 $1.84\times 10^0$ & $7.17$ & $1.07$ &
 $2.09\times 10^0$ & $6.81$ & $1.49$ \\
 &&& $10.0-25.0$ & $1.04\times 10^1$ & $16.24$ & $2.86$ &
 $1.01\times 10^1$ & $16.74$ & $2.95$ &
 $1.04\times 10^1$ & $16.54$ & $2.89$ &
 $1.04\times 10^1$ & $16.31$ & $2.84$ \\
 & $2$ & $0$ & $10.0-20.0$ & $4.09\times 10^3$ & $14.48$ & $1.52$ &
 $4.09\times 10^3$ & $14.38$ & $1.53$ &
 $4.18\times 10^3$ & $14.24$ & $1.56$ &
 $4.09\times 10^3$ & $14.42$ & $1.54$ \\
 && $1$ & $~3.0-12.8$ & $8.85\times 10^2$ & $9.06$ & $2.32$ &
 $8.72\times 10^2$ & $8.98$ & $2.25$ &
 $8.78\times 10^2$ & $10.80$ & $1.43$ &
 $8.92\times 10^2$ & $8.98$ & $2.16$ \\
 &&& $12.8-20.0$ & $1.28\times 10^3$ & $15.16$ & $1.54$ &
 $1.31\times 10^3$ & $15.01$ & $1.47$ &
 $1.11\times 10^3$ & $15.25$ & $1.44$ &
 $1.30\times 10^3$ & $15.30$ & $1.43$ \\
 & $3$ & $0$ & $~3.5-15.0$ & $1.30\times 10^5$ & $7.97$ & $2.72$ &
 $1.31\times 10^5$ & $7.91$ & $2.73$ &
 $8.66\times 10^4$ & $7.89$ & $2.66$ &
 $1.34\times 10^5$ & $7.94$ & $2.67$ \\
 &&& $15.0-30.0$ & $8.49\times 10^4$ & $23.78$ & $4.00$ &
 $8.35\times 10^4$ & $23.74$ & $3.91$ &
 $9.84\times 10^4$ & $22.49$ & $3.85$ &
 $8.91\times 10^4$ & $24.09$ & $3.95$ \\
 & $3$ & $1$ & $~3.5-15.0$ & $1.31\times 10^5$ & $8.47$ & $2.86$ &
 $1.28\times 10^5$ & $8.48$ & $2.89$ &
 $8.15\times 10^4$ & $9.08$ & $2.98$ &
 $1.32\times 10^5$ & $8.43$ & $2.80$ \\
 &&& $15.0-30.0$ & $6.13\times 10^4$ & $22.81$ & $4.15$ &
 $6.13\times 10^4$ & $22.88$ & $4.12$ &
 $7.73\times 10^4$ & $21.73$ & $4.11$ &
 $6.40\times 10^4$ & $23.07$ & $4.19$ \\
\hline\hline
%\end{tabular}
%\end{center}
\end{longtable}
\end{small}
\end{minipage}}

In Table~\ref{tab:summed}
we find that the errors in the GEM results are
$0.3\,\mathrm{MeV}$ for $\bar{\omega}^{(\lambda,\tau)}_D$,
$5\%$ for $B^{(\lambda,\tau)}_D$
and $0.2\,\mathrm{MeV}$ for $\sigma^{(\lambda,\tau)}_D$ at maximum,
which are compared to $0.5\,\mathrm{MeV}$ ($1.1\,\mathrm{MeV}$),
$10\%$ ($30\%$) and $0.1\,\mathrm{MeV}$ ($0.9\,\mathrm{MeV}$)
in the box-boundary (the quasi-HO) results.
We thus confirm that the GEM reproduces distribution
of the transition strengths with good precision
up to $\sigma^{(\lambda,\tau)}_D$,
from stable to drip-line nuclei.

\begin{figure}
%\centerline{\includegraphics[scale=0.85]{delta/trd_Ca60E3low.eps}}
\centerline{\includegraphics[scale=0.85]{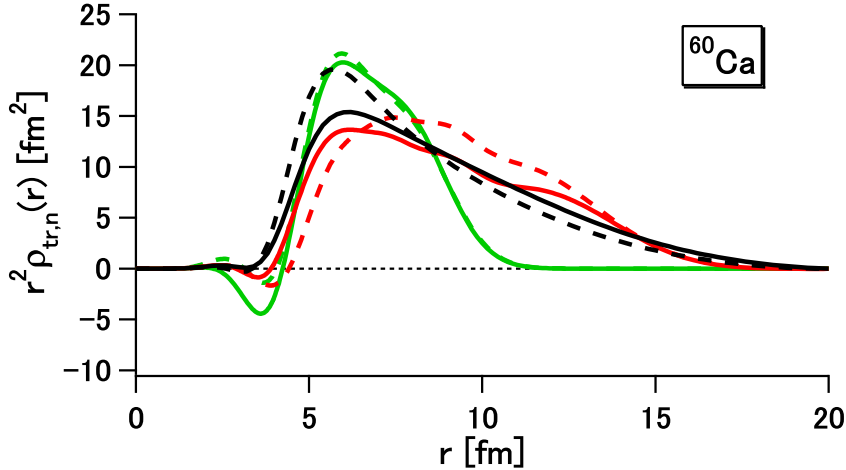}}
\vspace{2mm}
\caption{$r^2\rho_{\mathrm{tr},n}^{(\lambda=3)}(r;\alpha)$
 for low-energy states $\alpha$ in $^{60}$Ca.
 The red solid (dashed) line is obtained by the GEM
 at $\omega_\alpha=5.02\,\mathrm{MeV}$ ($6.68\,\mathrm{MeV}$),
 while the green solid (dashed) line by the quasi-HO basis functions
 at $\omega_\alpha=5.11\,\mathrm{MeV}$ ($7.37\,\mathrm{MeV}$).
 All these states carry relatively strong transition strengths
 of $B^{(\lambda=3,\tau=1)}_\alpha>5\times 10^3\,\mathrm{fm}^6$.
 The black solid (dashed) line is the continuum RPA result
 at $\omega=5.02\,\mathrm{MeV}$ ($6.92\,\mathrm{MeV}$).
\label{fig:trd_Ca60E3}}
\end{figure}

In Figs.~\ref{fig:E1IV_comp}--\ref{fig:E3IV_comp},
we view considerable transition strengths
at low $\omega$ in $^{60}$Ca,
which arise due to the loosely bound neutrons.
Similar results have been reported in the continuum RPA calculations
with the Skyrme interaction~\cite{ref:HSZ97,ref:HSZ01,ref:Sag01}.
Whereas such low-lying transition strengths in drip-line nuclei
will be investigated further in a forthcoming paper,
we here argue dependence of the results on the computational methods.
In comparison to the results obtained with the other methods,
notable deviation is found in the quasi-HO results
for the strengths at $\omega\lesssim 10\,\mathrm{MeV}$.
This is because the quasi-HO basis functions are not suitable
for describing loosely bound nucleons
which may have broad spatial distribution.
To clarify this point,
we consider the transition density defined by
\begin{equation}
 r^2\rho_{\mathrm{tr},\tau_z}^{(\lambda)}(r;\alpha)
  = \langle \alpha|\sum_{i\in\tau_z} \delta(r-r_i)\,
  r_i^\lambda Y^{(\lambda)}(\hat{\mathbf{r}}_i)|0\rangle
  \quad(\tau_z=p,n)\,.
 \label{eq:trdns}
\end{equation}
In Fig.~\ref{fig:trd_Ca60E3},
the neutron transition densities are depicted
for the low-energy $\lambda=3$ modes of $^{60}$Ca.
We present the transition densities at two prominent peaks
in $5<\omega<7.5\,\mathrm{MeV}$
obtained in the GEM and the quasi-HO calculations.
For comparison, the densities in the continuum RPA are also displayed,
whose renormalization factors ($C$ in Eq.~(17) of Ref.~\cite{ref:MMS05})
are determined so that the densities should be comparable
to their counterparts in the GEM and the quasi-HO results.

It is found that the transition densities depend
primarily on the s.p. bases, \textit{i.e.} the GEM or the quasi-HO,
rather than on the excitation energies of the states.
These low-energy transitions are dominated by excitation of a neutron
from the $pf$-shell to the continuum.
The rapid decrease of $r^2\rho_{\mathrm{tr},n}^{(\lambda=3)}(r;\alpha)$
at $r\approx 9\,\mathrm{fm}$ in the quasi-HO results
is attributed to limitation of the basis functions
mentioned above.
It does not seem easy to describe such transitions properly,
even if we increase the number of the quasi-HO basis functions.
On the contrary, the GEM gives broad distribution
of $r^2\rho_{\mathrm{tr},n}^{(\lambda=3)}(r;\alpha)$,
in fair agreement with the corresponding transition densities
in the continuum RPA.
This implies that by the GEM we can take into account
the effects of coupling to the continuum sufficiently.
We here comment that $r^2\rho_{\mathrm{tr},n}^{(\lambda=3)}(r;\alpha)$
obtained with the box-boundary method
is also close to that of the continuum RPA in this energy region.

We have thus established that the method
using the $A$-independent GEM basis functions of Eq.~(\ref{eq:basis-param})
is capable of describing the excitation energies, the transition strengths
and the widths of their distribution with good precision,
including nuclei in vicinity of the drip line,
within the RPA framework.
This method gives precision comparable to
the method assuming the box boundary,
and substantially better than the method using the quasi-HO basis functions
for nuclei near the neutron drip line.
The present method based on the GEM
cannot reproduce the smooth energy dependence
of the transition strengths in the continuum region.
This is a defect if compared to the continuum RPA method.
On the other hand, the GEM has an advantage in adaptability
to finite-range interactions.
In the next section
we shall apply the present method to self-consistent RPA calculations
with a finite-range interaction.

\section{Application of GEM to self-consistent RPA
 \label{sec:scRPA}}

The numerical method based on the GEM is applied
to self-consistent RPA calculations.
We consider the effective nuclear Hamiltonian
comprised of the kinetic energy and the effective $NN$ interaction,
\begin{equation}
 H_N = K + V_N\,;\quad
 K = \sum_i \frac{\mathbf{p}_i^2}{2M}\,,\quad
 V_N = \sum_{i<j} v_{ij}\,.
\label{eq:Hamil}\end{equation}
Here $i$ and $j$ are indices of the constituent nucleons.
The full Hamiltonian is given by $H=H_N+V_C-H_\mathrm{c.m.}$,
where $V_C$ stands for the Coulomb interaction among protons
and $H_\mathrm{c.m.}$ is the c.m. Hamiltonian.
We use the $A$-independent GEM basis functions
of Eq.~(\ref{eq:basis-param})
for all the calculations in this section.
Once we compute and store the two-body matrix elements of $V_N$,
we can use them both for the MF and the RPA calculations
(except the density-dependent part of $V_N$).
The exchange term of $V_C$ is exactly treated~\cite{ref:NS02}.
It should be commented that $H_\mathrm{c.m.}$ is highly non-local,
and therefore is not easy to be handled in the coordinate representation.
In contrast, we can easily take into account
both the one- and two-body terms of $H_\mathrm{c.m.}$ in the GEM.

In the calculations shown below,
we adopt the D1S parameter-set~\cite{ref:D1S}
of the Gogny interaction~\cite{ref:Gogny} for $V_N$,
in which the central channel has finite ranges,
and take $^{40,48,60}$Ca as examples.
Note that $^{60}$Ca is located near the neutron drip line
in the prediction with D1S~\cite{ref:Nak08b}.
We carry out the RPA calculations on top of the spherical HF solutions
for these nuclei.

In Ref.~\cite{ref:PG08},
the D1S interaction was applied to the quasiparticle RPA calculations
for the Si and Mg nuclei near the $\beta$-stability line,
by employing the HO basis functions.
The present method based on the GEM will be useful
for extending such approaches to drip-line nuclei.

\subsection{Spurious c.m. motion\label{subsec:spurious}}

It is proved that the spurious states,
which are the Nambu-Goldstone modes
emerging from the spontaneous symmetry breaking,
have zero excitation energy
and are separated from the other states
in the self-consistent RPA~\cite{ref:RS80}.
A typical example is the spurious c.m. motion.
However, we do not necessarily have a zero-energy state
in practical calculations
even if the Hamiltonian keeps the translational invariance,
because of finite size of the s.p. space.
Conversely, the energy of the spurious c.m. state provides a measure
of accuracy of the numerical calculation.

We here express the lowest $1^-$ state
in the solution of the RPA equation,
which corresponds to the spurious c.m. motion, by $\alpha=s$.
In Table~\ref{tab:cm}, $\omega_s^2$ values
obtained by the GEM basis-set of Eq.~(\ref{eq:basis-param})
are shown for $^{40,48,60}$Ca.
By the GEM the spurious state has zero energy to good precision.
Moreover, if the spurious state is well separated,
the transition strength $B_\alpha^{(\lambda=1,\tau=0)}$
should vanish for $\alpha\ne s$.
We confirm $B_\alpha^{(\lambda=1,\tau=0)}<10^{-3}\,\mathrm{fm}^2$
for all $\alpha(\ne s)$, in any of the three nuclei.

\begin{table}\begin{center}
\caption{$\omega_s^2$ values ($\mathrm{MeV}^2$)
 in the self-consistent RPA calculation by the GEM
 with the D1S interaction.
\label{tab:cm}}
\begin{tabular}{cr}
\hline\hline
nuclide & \multicolumn{1}{c}{$\omega_s^2$} \\
 \hline
$^{40}$Ca & $-5.80\times 10^{-6}$ \\
$^{48}$Ca & $-8.61\times 10^{-6}$ \\
$^{60}$Ca & $-2.67\times 10^{-6}$ \\
\hline\hline
\end{tabular}
\end{center}\end{table}

\subsection{Energy-weighted sum rules\label{subsec:EWSR}}

The RPA preserves the energy-weighted sum rule (EWSR),
\begin{equation}
 \Sigma_1 = \sum_\alpha \omega_\alpha\,
  \big|\langle\alpha|\mathcal{O}|0\rangle\big|^2
  = \frac{1}{2}\langle 0|[\mathcal{O}^\dagger,[H,\mathcal{O}]]
  |0\rangle\,,
 \label{eq:EWSR}
\end{equation}
if the expectation value of the double commutator
is evaluated for the MF ground state~\cite{ref:RS80}.
In correspondence to the specific mode $\mathcal{O}^{(\lambda,\tau)}$,
we define $\Sigma_1^{(\lambda,\tau)}$.
The energy-weighted sum $\Sigma_1^{(\lambda,\tau)}$
is associated with the quantities in Eq.~(\ref{eq:B-dist})
by $\Sigma_1^{(\lambda,\tau)}=\bar{\omega}^{(\lambda,\tau)}_D
\cdot B^{(\lambda,\tau)}_D$,
for $D$ covering the whole range of $\omega$
(\textit{i.e.} $\omega_D^\mathrm{min}=0$
and $\omega_D^\mathrm{max}=\infty$).
In evaluating the double commutator,
we first take $H'=H_N+V_C$, ignoring $H_\mathrm{c.m.}$,
instead of the full Hamiltonian $H$.
Most effective interactions (or energy density functionals)
including the Skyrme, the Gogny
and the M3Y-type ones~\cite{ref:Nak03,ref:Nak08b}
lead to the continuity equation for the isoscalar density and current
(as long as we use $H'$),
from which the following EWSR for the isoscalar transition
is derived~\cite{ref:Suz80},
\begin{eqnarray}
 \Sigma_1^{\prime\,(\lambda,\tau=0)} &=& \frac{1}{2}
  \langle 0|[\mathcal{O}^{(\lambda,\tau=0)\dagger},
  [H',\mathcal{O}^{(\lambda,\tau=0)}]]|0\rangle \nonumber\\
 &=& \frac{1}{2} \langle 0|[\mathcal{O}^{(\lambda,\tau=0)\dagger},
  [K,\mathcal{O}^{(\lambda,\tau=0)}]]|0\rangle
 ~=~ \frac{\lambda(2\lambda+1)^2}{4\pi}\,\frac{1}{2M}
 \langle 0|\sum_i r_i^{2\lambda-2}|0\rangle\,,
 \label{eq:cSR1}
\end{eqnarray}
where we restrict ourselves to $\lambda\geq 2$.
Though not written explicitly,
the $z$-components of $\mathcal{O}^{(\lambda,\tau)}$ are summed up.
However, $H_\mathrm{c.m.}$ modifies the EWSR as
\begin{eqnarray}
 \Sigma_1^{(\lambda,\tau=0)} &=& \frac{1}{2}
  \langle 0|[\mathcal{O}^{(\lambda,\tau=0)\dagger},
  [H,\mathcal{O}^{(\lambda,\tau=0)}]]|0\rangle \nonumber\\
 &=& \frac{\lambda(2\lambda+1)^2}{4\pi}\,\frac{1}{2M} \bigg[
 \langle 0|\sum_i r_i^{2\lambda-2}|0\rangle \nonumber\\
 &&\qquad - \frac{4\pi}{2\lambda-1}\,\frac{1}{A}
 \langle 0|\big(\sum_i r_i^{\lambda-1}
 Y^{(\lambda-1)}(\hat{\mathbf{r}}_i)\big)\cdot\big(\sum_i r_i^{\lambda-1}
 Y^{(\lambda-1)}(\hat{\mathbf{r}}_i)\big)|0\rangle \bigg] \nonumber\\
 &=& \frac{\lambda(2\lambda+1)^2}{4\pi}\,\frac{1}{2M} \bigg[
 \big(1-\frac{1}{A}\big)\langle 0|\sum_i r_i^{2\lambda-2}|0\rangle
 \nonumber\\
 &&\qquad - \frac{4\pi}{2\lambda-1}\,\frac{1}{A}
 \langle 0|\sum_{i\ne j}
 r_i^{\lambda-1} Y^{(\lambda-1)}(\hat{\mathbf{r}}_i)\cdot
 r_j^{\lambda-1} Y^{(\lambda-1)}(\hat{\mathbf{r}}_j)|0\rangle \bigg]\,.
 \label{eq:cSR2}
\end{eqnarray}
For the $\lambda=2$ case,
Eq.~(\ref{eq:cSR2}) can be rewritten as
\begin{equation}
 \Sigma_1^{(\lambda=2,\tau=0)} =
 \frac{50}{4\pi}\,\frac{1}{2M}\,
 \langle 0|\sum_i (\mathbf{r}_i-\mathbf{R})^2|0\rangle\,,
 \label{eq:cSR3}
\end{equation}
where $\mathbf{R}$ denotes the c.m. position.
This correction due to $H_\mathrm{c.m.}$ is harmonious
with the c.m. correction to the rms matter radius~\cite{ref:Nak03}.
The EWSR of Eq.~(\ref{eq:cSR2}) can be used as another tool
to check appropriateness of the s.p. space
adopted in the numerical calculation.

For the $(\lambda=1,\tau=1)$ mode,
the Thomas-Reich-Kuhn (TRK) sum rule is derived if $V_N$ is local,
\begin{equation}
 \Sigma_\mathrm{TRK} = \frac{1}{2}
  \langle 0|[\tilde{\mathcal{O}}^{(\lambda=1,\tau=1)\dagger},
  [K,\tilde{\mathcal{O}}^{(\lambda=1,\tau=1)}]]|0\rangle
 = \frac{9}{4\pi}\,\frac{1}{2M}\,\frac{4ZN}{A}\,.
 \label{eq:TRK}
\end{equation}
However, $\Sigma_1^{(\lambda=1,\tau=1)}$ shifts
from $\Sigma_\mathrm{TRK}$ in practice, owing to the non-locality
in the charge exchange term of $V_N$.
The enhancement factor $\kappa$ is defined by
\begin{equation}
 \Sigma_1^{(\lambda=1,\tau=1)}
  = (1+\kappa)\,\Sigma_\mathrm{TRK}\,.
 \label{eq:kappa}
\end{equation}
It is noted that $H_\mathrm{c.m.}$ does not influence
the energy-weighted sum of the $(\lambda=1,\tau=1)$ mode
(\textit{i.e.} $[\tilde{O}^{(\lambda=1,\tau=1)\dagger},
[H_\mathrm{c.m.},\tilde{O}^{(\lambda=1,\tau=1)}]]=0$),
as long as it is fully taken into account.
The enhancement factor $\kappa$ can be expressed
by an expectation value of the charge exchange term of $V_N$
for the ground state.
In the case of the Skyrme interaction,
an explicit expression of $\kappa$
in terms of the neutron and proton densities
has been given in Ref.~\cite{ref:SSAR}.
It is complicated to evaluate $\kappa$ in an individual nucleus
if $V_N$ is taken to be a finite-range interaction.
However, $\kappa$ in the symmetric nuclear matter
is related to the Landau-Migdal parameter $f'_1$~\cite{ref:Nak03} by
\begin{equation}
 1+\kappa = \frac{M}{M^\ast}\big(1+\frac{f'_1}{3})\,,
 \label{eq:f'1}
\end{equation}
where $M^\ast$ denotes the effective $k$-mass.
Equation~(\ref{eq:f'1}) gives an approximate value of $\kappa$
for finite nuclei.

In many calculations using the Skyrme interaction,
only the one-body term of $H_\mathrm{c.m.}$ is taken into account.
We point out that this prescription influences the above sum rules.
For the $(\lambda\geq 2, \tau=0)$ modes,
the energy-weighted sum becomes $\Sigma_1^{\prime\,(\lambda,\tau=0)}$
multiplied by $(1-1/A)$,
lacking the second term in the last expression of Eq.~(\ref{eq:cSR2}).
Moreover, if we subtract only the one-body term of $H_\mathrm{c.m.}$,
the TRK part of the energy-weighted sum for the $(\lambda=1,\tau=1)$ mode
(\textit{i.e.} Eq.~(\ref{eq:TRK})) is fictitiously reduced by $(1-1/A)$.

As an indicator for the EWSR, we define the ratio,
\begin{equation}
 \mathcal{R}^{(\lambda,\tau)}
  = \sum_\alpha \omega_\alpha\,
  \big|\langle\alpha|\mathcal{O}^{(\lambda,\tau)}|0\rangle\big|^2
 \bigg/ \frac{1}{2}\langle 0|[\mathcal{O}^{(\lambda,\tau)\dagger},
  [H,\mathcal{O}^{(\lambda,\tau)}]]|0\rangle\,,
\end{equation}
which should be unity if the s.p. basis-set is complete.
In Table~\ref{tab:cSR}, we show $\mathcal{R}^{(\lambda,\tau=0)}$
($\lambda=2,3$) in $^{40,48,60}$Ca.
It is confirmed that the present method satisfies
the EWSRs of Eq.~(\ref{eq:cSR2}) within a few percent precision.
For the $(\lambda=1,\tau=1)$ mode,
we use the ratio to the TRK sum rule,
\begin{equation}
 \tilde{\mathcal{R}}_K^{(\lambda=1,\tau=1)}
  = \sum_\alpha \omega_\alpha\,
  \big|\langle\alpha|\tilde{\mathcal{O}}^{(\lambda=1,\tau=1)}
  |0\rangle\big|^2 \bigg/
  \frac{1}{2}\langle 0|[\tilde{\mathcal{O}}^{(\lambda=1,\tau=1)\dagger},
  [K,\tilde{\mathcal{O}}^{(\lambda=1,\tau=1)}]]|0\rangle\,,
\end{equation}
and compare it with the rhs of Eq.~(\ref{eq:f'1}),
which is denoted by
$\tilde{\mathcal{R}}_{K,\infty}^{(\lambda=1,\tau=1)}$.
The values of $\tilde{\mathcal{R}}_K^{(\lambda=1,\tau=1)}(=1+\kappa)$
are presented in Table~\ref{tab:cSR1}.
The $\tilde{\mathcal{R}}_K^{(\lambda=1,\tau=1)}$ value
does not depend strongly on nuclides.
The nuclear matter value of
$\tilde{\mathcal{R}}_{K,\infty}^{(\lambda=1,\tau=1)}$
gives a good first approximation,
and slight reduction
from $\tilde{\mathcal{R}}_{K,\infty}^{(\lambda=1,\tau=1)}$
may be accounted for mainly by the difference
between $\rho_p(\mathbf{r})$ and $\rho_n(\mathbf{r})$~\cite{ref:SSAR}.
It is noted that the experimental value has been reported as
$\tilde{\mathcal{R}}_K^{(\lambda=1,\tau=1)}=1.76\pm 0.10$~\cite{ref:kappa},
being almost independent of the mass number.

\begin{table}\begin{center}
\caption{$\mathcal{R}^{(\lambda,\tau=0)}$
 in the RPA calculation by the GEM,
 in which the D1S interaction is employed.
\label{tab:cSR}}
\begin{tabular}{ccr}
\hline\hline
nuclide & $\lambda$
 & \multicolumn{1}{c}{$\mathcal{R}^{(\lambda,\tau=0)}$} \\
 \hline
$^{40}$Ca & $2$ & $1.005$ \\
 & $3$ & $1.031$ \\
$^{48}$Ca & $2$ & $1.006$ \\
 & $3$ & $1.033$ \\
$^{60}$Ca & $2$ & $1.003$ \\
 & $3$ & $1.010$ \\
\hline\hline
\end{tabular}
\end{center}\end{table}

\begin{table}\begin{center}
\caption{$\tilde{\mathcal{R}}_K^{(\lambda=1,\tau=1)}$
 in the RPA calculation by the GEM,
 in comparison with the nuclear matter value
 $\tilde{\mathcal{R}}_{K,\infty}^{(\lambda=1,\tau=1)}
 = (M/M^\ast)\,(1+f'_1/3)$.
 The D1S interaction is used.
\label{tab:cSR1}}
\begin{tabular}{crr}
\hline\hline
nuclide
 & \multicolumn{1}{c}{$\tilde{\mathcal{R}}_K^{(\lambda=1,\tau=1)}$}
 & \multicolumn{1}{c}
 {$\tilde{\mathcal{R}}_{K,\infty}^{(\lambda=1,\tau=1)}$} \\
 \hline
$^{40}$Ca & $1.587$ & \\
$^{48}$Ca & $1.589$ & $1.660$ \\
$^{60}$Ca & $1.584$ & \\
\hline\hline
\end{tabular}
\end{center}\end{table}

\subsection{Strength functions\label{subsec:S(E)}}

\begin{figure}
%\centerline{\includegraphics[scale=0.85]{D1S/E1IV_Ca_D1S.eps}}
\centerline{\includegraphics[scale=0.85]{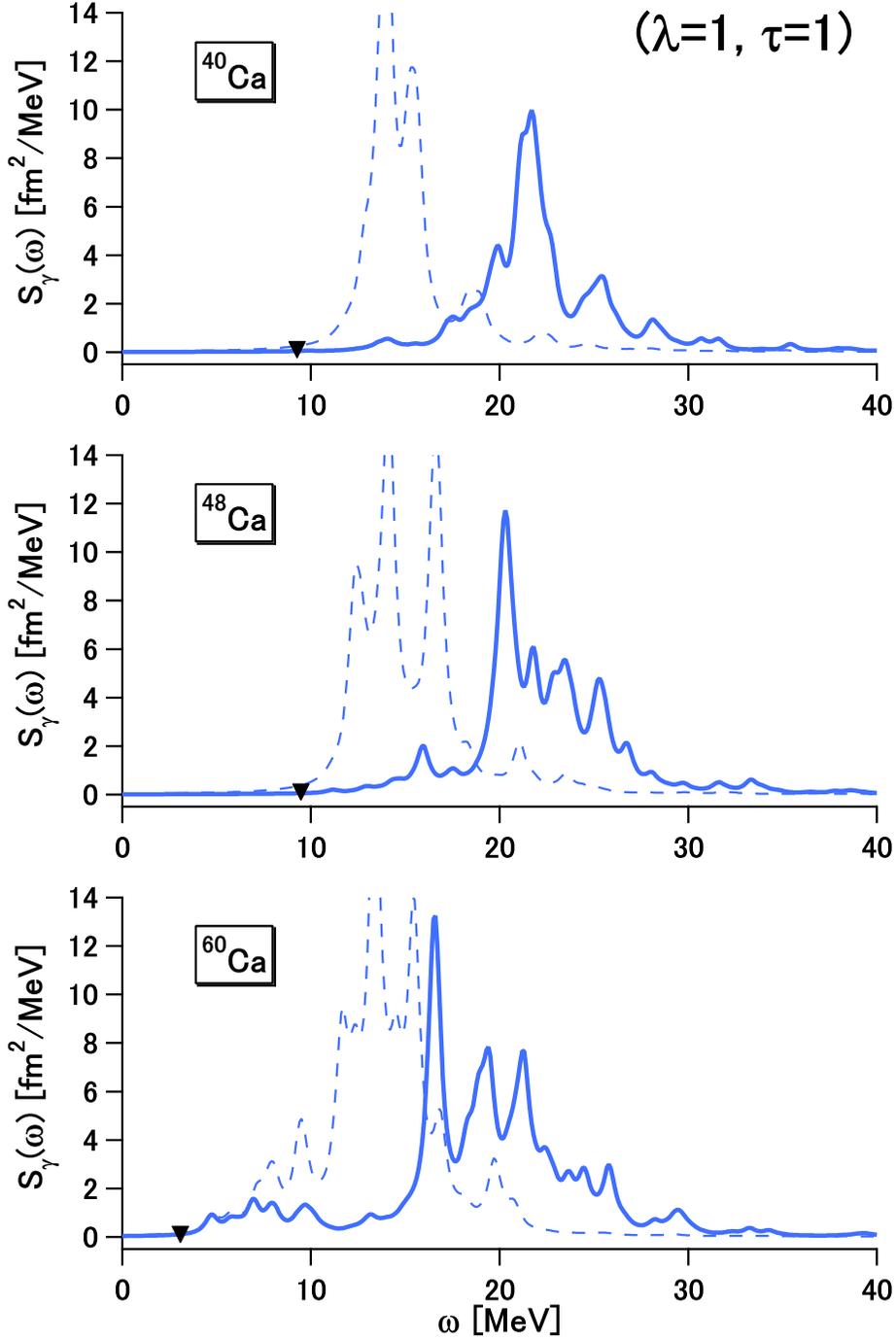}}
\vspace{2mm}
\caption{$\tilde{S}_\gamma^{(\lambda=1,\tau=1)}(\omega)$
 in $^{40,48,60}$Ca.
 We take $\gamma=0.4\,\mathrm{MeV}$.
 The solid lines are obtained by the self-consistent RPA calculations
 with the D1S interaction,
 while the dashed lines represent the unperturbed strength functions.
 The particle threshold energy in the HF approximation
 is indicated by the inverted triangle for each nucleus.
\label{fig:E1IV_D1S}}
\end{figure}

\begin{figure}
%\centerline{\includegraphics[scale=0.85]{D1S/E2IS_Ca_D1S.eps}}
\centerline{\includegraphics[scale=0.85]{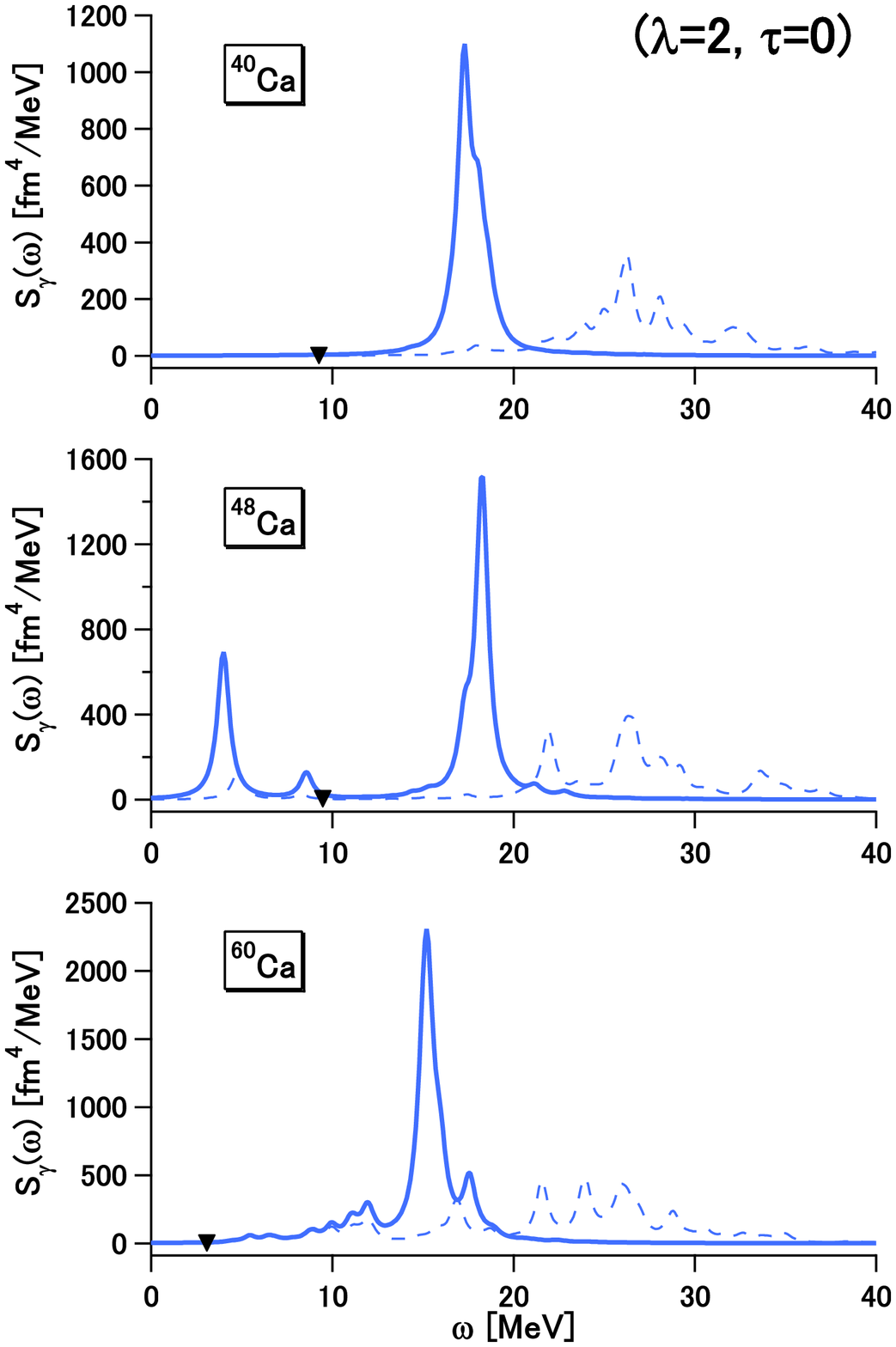}}
\vspace{2mm}
\caption{$S_\gamma^{(\lambda=2,\tau=0)}(\omega)$
 in $^{40,48,60}$Ca calculated with the D1S interaction.
 Conventions are the same as in Fig.~\protect\ref{fig:E1IV_D1S}.
\label{fig:E2IS_D1S}}
\end{figure}

\begin{figure}
%\centerline{\includegraphics[scale=0.85]{D1S/E2IV_Ca_D1S.eps}}
\centerline{\includegraphics[scale=0.85]{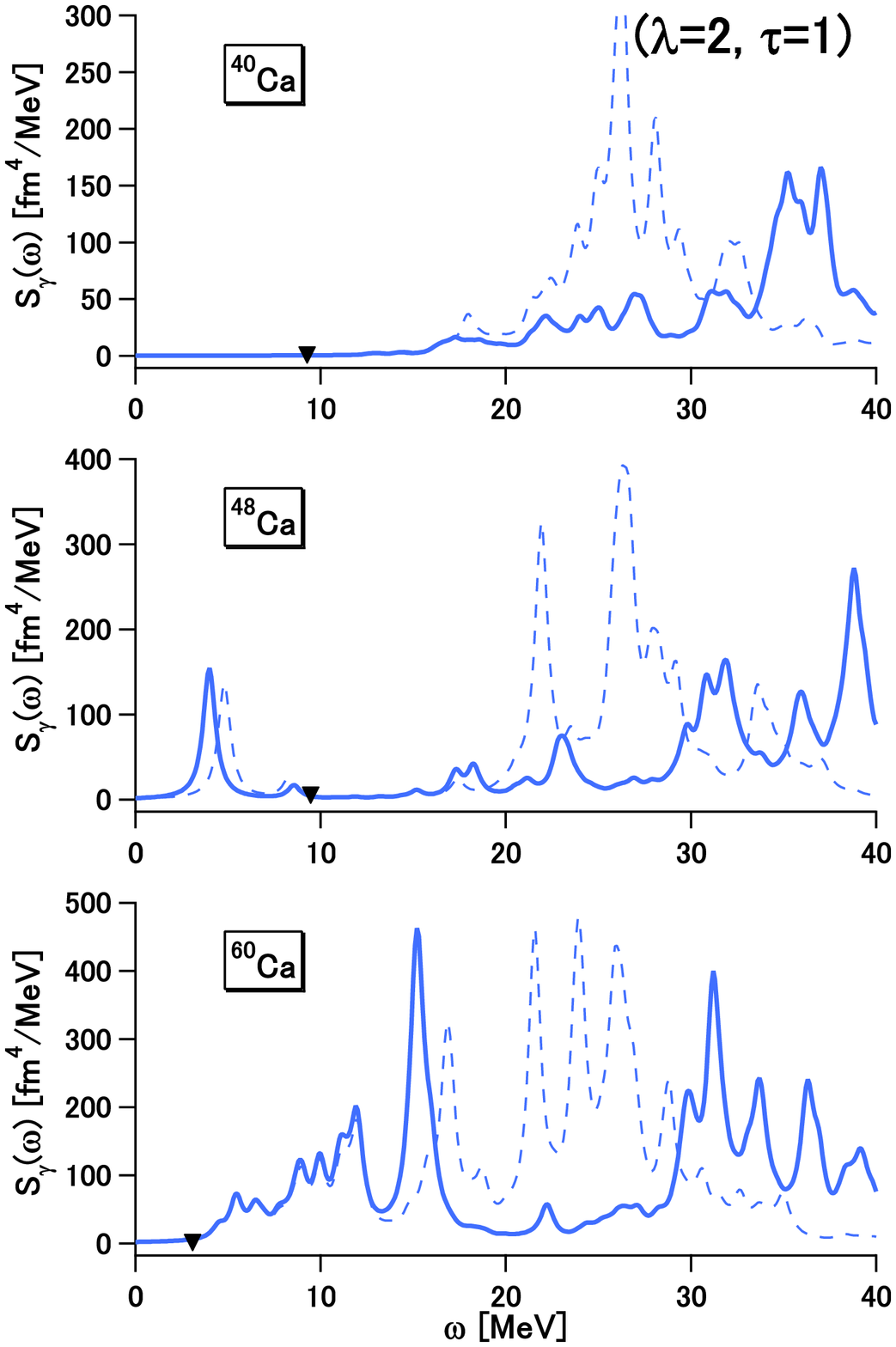}}
\vspace{2mm}
\caption{$S_\gamma^{(\lambda=2,\tau=1)}(\omega)$
 in $^{40,48,60}$Ca calculated with the D1S interaction.
 Conventions are the same as in Fig.~\protect\ref{fig:E1IV_D1S}.
\label{fig:E2IV_D1S}}
\end{figure}

\begin{figure}
%\centerline{\includegraphics[scale=0.85]{D1S/E3IS_Ca_D1S.eps}}
\centerline{\includegraphics[scale=0.85]{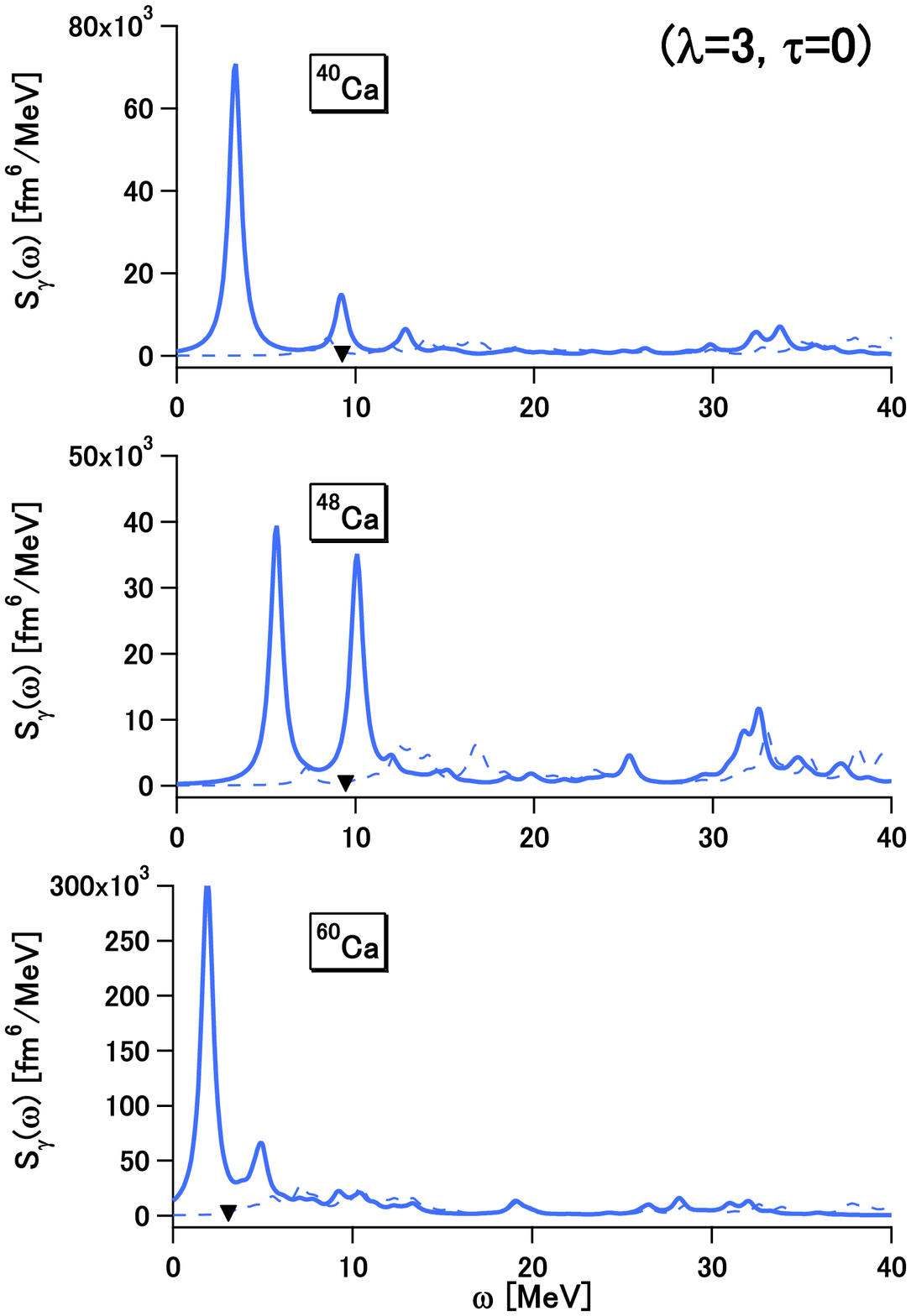}}
\vspace{2mm}
\caption{$S_\gamma^{(\lambda=3,\tau=0)}(\omega)$
 in $^{40,48,60}$Ca calculated with the D1S interaction.
 Conventions are the same as in Fig.~\protect\ref{fig:E1IV_D1S}.
\label{fig:E3IS_D1S}}
\end{figure}

\begin{figure}
%\centerline{\includegraphics[scale=0.85]{D1S/E3IV_Ca_D1S.eps}}
\centerline{\includegraphics[scale=0.85]{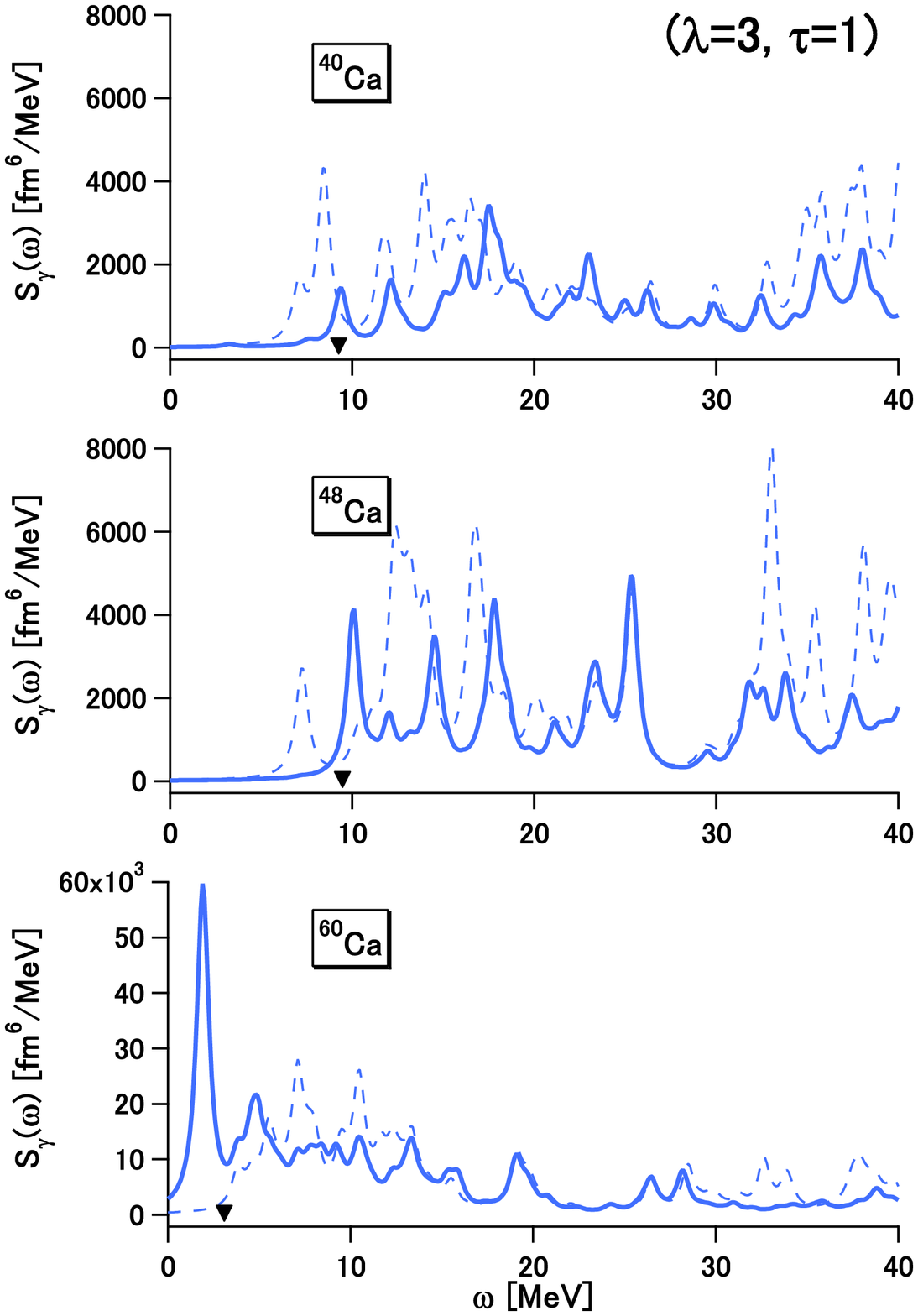}}
\vspace{2mm}
\caption{$S_\gamma^{(\lambda=3,\tau=1)}(\omega)$
 in $^{40,48,60}$Ca calculated with the D1S interaction.
 Conventions are the same as in Fig.~\protect\ref{fig:E1IV_D1S}.
\label{fig:E3IV_D1S}}
\end{figure}

We next show the strength functions $S_\gamma^{(\lambda,\tau)}(\omega)$
obtained by the self-consistent RPA calculations with D1S,
in Figs.~\ref{fig:E1IV_D1S}--\ref{fig:E3IV_D1S}.
We here adopt $\gamma=0.4\,\mathrm{MeV}$.
The strength functions in the RPA
are compared with the unperturbed strength functions,
in which the residual interaction is ignored.

Irrespectively of $(\lambda,\tau)$,
the strength functions $S_\gamma^{(\lambda,\tau)}(\omega)$
with the D1S force
are qualitatively similar to
those obtained with the schematic interaction
shown in Sec.~\ref{sec:contact}.
Significant difference between the RPA strength function
and the unperturbed one
suggests collectivity of the transition.
In this regard we view collectivity
of the isovector giant dipole resonance (IV-GDR)
induced by a repulsive part of the residual interaction
in Fig.~\ref{fig:E1IV_D1S},
and of the isoscalar giant quadrupole and octupole resonances
(IS-GQR and IS-GOR) induced by an attractive interaction
in Figs.~\ref{fig:E2IS_D1S} and \ref{fig:E3IS_D1S}.
Collectivity of these giant resonances is further confirmed
from the forward and backward amplitudes of the state
that forms the highest peak in the strength function;
a number of the unperturbed excitations are mixed
in the RPA states.

\begin{figure}
%\centerline{\includegraphics[scale=0.85]{D1S/E1td_Ca60_D1S.eps}}
\centerline{\includegraphics[scale=0.85]{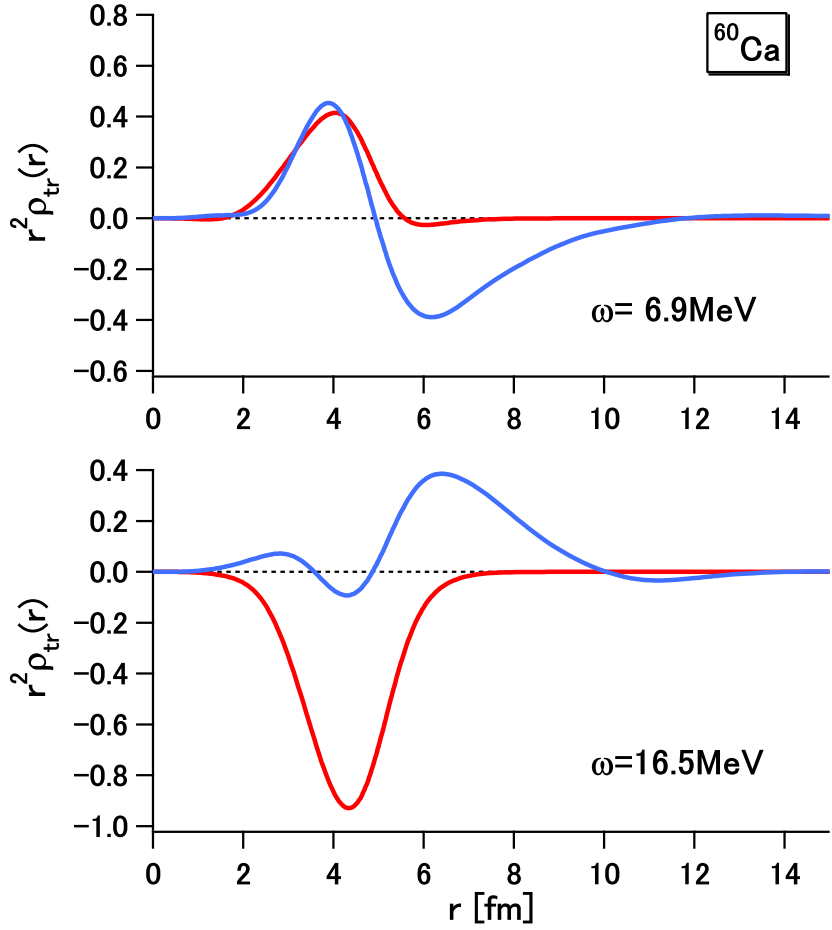}}
\vspace{2mm}
\caption{$r^2\tilde{\rho}_{\mathrm{tr},\tau_z}^{(\lambda=1)}(r;\alpha)$
 in $^{60}$Ca  for the $\omega_\alpha=6.9$ and $16.5\,\mathrm{MeV}$ states,
 obtained by the RPA calculation with the D1S interaction.
 The blue (red) line is for $\tau_z=n$ ($p$).
\label{fig:E1td_D1S}}
\end{figure}

In $^{60}$Ca,
transition strengths emerge and form a peak
in $S_\gamma^{(\lambda,\tau)}$
at relatively low energy ($\omega\lesssim 10\,\mathrm{MeV}$),
though the transitions are not so strong as in the giant resonances.
It is of interest whether such low-energy strengths have
collective nature or not.
By comparing the RPA strength functions to the unperturbed ones,
it is suggested that the low-energy transitions specific to $^{60}$Ca
hardly have strong collectivity.
This consequence is consistent with the argument in Ref.~\cite{ref:CDV96},
and will further be investigated in a forthcoming paper.
For the $(\lambda=1,\tau=1)$ mode,
the low-energy strengths may be compared
to the so-called pygmy dipole resonance (PDR)
that has been observed in several $Z<N$ nuclei
such as $^{90}$Zr~\cite{ref:PDR-Zr90} and $^{208}$Pb~\cite{ref:PDR-Pb208}.
The IV-GDR in $Z\sim N$ nuclei is interpreted
as a surface oscillation
to which the protons and the neutrons contribute with opposite phases.
In the neutron-rich nuclei the $E1$ transition could have
two collective components;
an oscillation between protons and neutrons,
and an oscillation between a core and excess neutrons.
The latter is a possible interpretation of the PDR.
In Fig.~\ref{fig:E1td_D1S}
the neutron and proton transition densities in $^{60}$Ca
are depicted for the $\omega_\alpha=6.9\,\mathrm{MeV}$
and $16.5\,\mathrm{MeV}$ states,
both of which form the peaks in Fig.~\ref{fig:E1IV_D1S}.
We here define the transition density by
\begin{equation}
 r^2\tilde{\rho}_{\mathrm{tr},\tau_z}^{(\lambda=1)}(r;\alpha)
  = \left\{\begin{array}{ll}
 {\displaystyle\frac{2Z}{A}\langle \alpha|\sum_{i\in n} \delta(r-r_i)\,
  r_i Y^{(1)}(\hat{\mathbf{r}}_i)|0\rangle} &
 (\mbox{for~}\tau_z=n)\\
 {\displaystyle\frac{2N}{A}\langle \alpha|\sum_{i\in p} \delta(r-r_i)\,
  r_i Y^{(1)}(\hat{\mathbf{r}}_i)|0\rangle} &
 (\mbox{for~}\tau_z=p) \end{array}
 \right.\,,
\end{equation}
which is analogous to Eq.~(\ref{eq:trdns})
but is subject to the modification of Eq.~(\ref{eq:op1}).
Because the spurious c.m. motion is well separated,
we have
\begin{equation}
 \int dr\cdot r^2 \rho_{\mathrm{tr},\mathrm{IS}}^{(\lambda=1)}
  (r;\alpha) = 0\,,
\label{eq:int-sptrd}
\end{equation}
where the isoscalar transition density is defined by
\begin{equation}
 r^2 \rho_{\mathrm{tr},\mathrm{IS}}^{(\lambda=1)}(r;\alpha)
 = \frac{A}{2NZ}\big[
 N\cdot r^2\tilde{\rho}_{\mathrm{tr},n}^{(\lambda=1)}(r;\alpha)
 + Z\cdot r^2\tilde{\rho}_{\mathrm{tr},p}^{(\lambda=1)}(r;\alpha)
 \big]\,.
\end{equation}
Figure~\ref{fig:E1td_D1S} shows that, to the first approximation,
the higher-lying strength corresponds to an out-of-phase oscillation
between protons and neutrons as in the usual IV-GDR,
while the lower-lying strength looks like an oscillation
between a core and outer neutrons,
as pointed out in Ref.~\cite{ref:TE06}.
It should be noted that the peaks
in $r^2\tilde{\rho}_{\mathrm{tr},\tau_z}^{(\lambda=1)}(r;\alpha)$
are considerably displaced between $\tau_z=n$ and $p$
for the transition to the $\omega_\alpha=16.5\,\mathrm{MeV}$ state.
This indicates that, whereas
$r^2 \rho_{\mathrm{tr},\mathrm{IS}}^{(\lambda=1)}(r;\alpha)$
is vanishingly small in the $Z\sim N$ nuclei such as $^{40}$Ca,
it does not vanish in $^{60}$Ca even for the strong transition,
though satisfying Eq.~(\ref{eq:int-sptrd}).
Moreover,
behavior of $r^2\tilde{\rho}_{\mathrm{tr},n}^{(\lambda=1)}(r;\alpha)$
at $r\approx 4\,\mathrm{fm}$ suggests that this state contains
a weak admixture of the oscillation
between a core and outer neutrons.

\section{Summary}
\label{sec:summary}

We have developed a method of implementing RPA calculations,
which is based on the Gaussian expansion method (GEM).
Owing to the advantages of the GEM
which have been established in the MF calculations,
it is naturally expected that we can efficiently compute
excitation of nuclei including coupling to the continuum.
The parameters of the s.p. basis functions are insensitive to nuclide,
and even calculations with a single set of bases
may cover wide range of the mass table.

The method has first been tested in $^{40,48,60}$Ca
with a density-dependent contact interaction
on top of the Woods-Saxon single-particle (s.p.) states,
by comparing its results with the results
obtained with the continuum RPA method.
We have confirmed that the present method
using the GEM basis functions of Eq.~(\ref{eq:basis-param})
describes the energies, the transition strengths
and the widths of their distribution with good precision
for the $1^-$, $2^+$ and $3^-$ collective states,
including drip-line nuclei,
although we cannot reproduce the smooth energy dependence
of the transition strengths in the continuum.
The new method has been compared also with the box-boundary method
and with the method using bases
analogous to the harmonic-oscillator (HO) ones.
The present method attains precision
comparable to the box-boundary method of $r_\mathrm{max}=20\,\mathrm{fm}$.
The basis-set similar to the HO set of $N_\mathrm{osc}\leq 11$
does not give good precision, particularly for low-energy transitions
in drip-line nuclei,
because it is unable to reproduce transition densities
at $r\gtrsim 10\,\mathrm{fm}$.
It may not be easy to describe such transitions appropriately
even if we take a larger number of the HO basis functions.

Another advantage of the present method
is its tractability of finite-range interactions.
This point has been demonstrated by the application
to the self-consistent HF plus RPA calculations in $^{40,48,60}$Ca
with the Gogny D1S interaction.
It has been confirmed that
the zero excitation energy of the spurious state
and the energy-weighted sum rules for the isoscalar transitions
are fulfilled to good precision.
By comparing the RPA strength functions to the unperturbed ones,
collectivity of the transitions has been argued.
For $^{60}$Ca, characters of the low-energy dipole strengths
have been investigated as well as those of the giant dipole resonance,
via the transition densities.

Further application of the present method to excitations of nuclei
with finite-range interactions,
including the semi-realistic interactions~\cite{ref:Nak03,ref:Nak08b},
is under progress.
We here mention a study of the $M1$ transition
in $^{208}$Pb~\cite{ref:Shi08},
in which significant effects of the tensor force have been confirmed.
Future plans include extension of the present method
to the quasiparticle RPA,
which is quite promising
since the GEM has been successfully applied
to the Hartree-Fock-Bogolyubov calculations~\cite{ref:Nak06}.
\\

\noindent
The authors are grateful to T. Nakatsukasa for discussions.
This work is financially supported in part
as Grant-in-Aid for Scientific Research (C), No.~19540262,
by Japan Society for the Promotion of Science.
Numerical calculations were performed on HITAC SR11000
at Institute of Media and Information Technology, Chiba University,
on HITAC SR11000 at Information Initiative Center, Hokkaido University,
and on NEC SX-8 at Yukawa Institute for Theoretical Physics,
Kyoto University.

\end{document}